\documentclass[conference,anonymous]{IEEEtran}

\usepackage{xspace}
\usepackage{listings}
\usepackage{xcolor}
\usepackage{graphicx}

\definecolor{mygreen}{rgb}{0,0.6,0}
\definecolor{mygray}{rgb}{0.5,0.5,0.5}
\definecolor{mymauve}{rgb}{0.58,0,0.82}
\definecolor{myorange}{rgb}{0.9,0.60,0}

\lstdefinelanguage
   [x64]{Assembler}     
   [x86masm]{Assembler} 
   {morekeywords={CDQE,CQO,CMPSQ,CMPXCHG16B,JRCXZ,LODSQ,MOVSXD, %
                  POPFQ,PUSHFQ,SCASQ,STOSQ,IRETQ,RDTSCP,SWAPGS, %
                  rax,rdx,rcx,rbx,rsi,rdi,rsp,rbp, %
                  r8,r8d,r8w,r8b,r9,r9d,r9w,r9b, %
                  r10,r10d,r10w,r10b,r11,r11d,r11w,r11b, %
                  r12,r12d,r12w,r12b,r13,r13d,r13w,r13b, %
                  r14,r14d,r14w,r14b,r15,r15d,r15w,r15b}} 

\newif\ifdebug
\debugfalse

\newcommand{\jelena}[1]{\ifdebug{\textcolor{blue}{\{jelena: #1\}}}\fi}

\newcommand{\mclearpage}{}

\newcommand{\modelextensionf}{\ensuremath{\mathcal{F}}\xspace}
\newcommand{\modelextensioncallee}{\ensuremath{\mathcal{C}_{1}}\xspace}
\newcommand{\modelextensionreturn}{\ensuremath{\mathcal{C}_{2}}\xspace}

\newcommand{\angrc}{angr\textsubscript{\tiny \ensuremath{\mathcal{C}}}\xspace}
\newcommand{\angrf}{angr\textsubscript{\tiny \ensuremath{\mathcal{F}}}\xspace}
\newcommand{\angrcf}{angr\textsubscript{\tiny \ensuremath{\mathcal{CF}}}\xspace}

\newcommand{\fscore}{F\textsubscript{1}\xspace}

\newcommand{\nummbsrcs}{$215,072$\xspace}
\newcommand{\nummbfuncs}{$277,072$\xspace}

\begin{document}

\title{Data Flows in You: Benchmarking and Improving Static Data-flow Analysis on Binary Executables}

\author{
    \IEEEauthorblockN{
        Nicolaas Weideman\IEEEauthorrefmark{1},
        Sima Arasteh\IEEEauthorrefmark{2},
        Mukund Raghothaman\IEEEauthorrefmark{2},
        Jelena Mirkovic\IEEEauthorrefmark{3},
        Christophe Hauser\IEEEauthorrefmark{4}
    }

    \IEEEauthorblockA{\IEEEauthorrefmark{1}USC Information Sciences Institute, Los Angeles, CA, USA\\
    nhweideman@gmail.com}

    \IEEEauthorblockA{\IEEEauthorrefmark{2}University of Southern California, Los Angeles, CA, USA\\
    \{arasteh, raghotha\}@usc.edu}

    \IEEEauthorblockA{\IEEEauthorrefmark{3}USC Information Sciences Institute, Los Angeles, CA, USA\\
    mirkovic@isi.edu}

    \IEEEauthorblockA{\IEEEauthorrefmark{4}Dartmouth College, Hanover, NH, USA\\
    christophe.hauser@dartmouth.edu}
}

\maketitle

\begin{abstract}
  Data-flow analysis is a critical component of security research. Theoretically, accurate data-flow analysis in binary executables is an undecidable problem, due to complexities of binary code. Practically, many binary analysis engines offer some data-flow analysis capability, but we lack understanding of the accuracy of these analyses, and their limitations.
  We address this problem by introducing a labeled benchmark data set, including \nummbsrcs microbenchmark test cases, mapping to \nummbfuncs binary executables, created specifically to evaluate data-flow analysis implementations.
  Additionally, we augment our benchmark set with dynamically-discovered data flows from $6$ real-world executables.
  Using our benchmark data set, we evaluate three state of the art data-flow analysis implementations, in angr, Ghidra and Miasm and discuss their very low accuracy and reasons behind it. We further propose three model extensions to static data-flow analysis that significantly improve accuracy, achieving almost perfect recall (0.99) and increasing precision from 0.13 to 0.32.
  Finally, we show that leveraging these model extensions in a vulnerability-discovery context leads to a tangible improvement in vulnerable instruction identification.
\end{abstract} \mclearpage

\IEEEpeerreviewmaketitle

\section{Introduction}
With the ever-increasing reliance of the modern world on software systems, the cat-and-mouse game of finding and mitigating software vulnerabilities before they are exploited has reached critical levels.
This is exacerbated by the growing complexity of software, which increases the burden of ensuring its security.
To assuage this burden,  program analysis for system security has gained more traction.
Data-flow analysis is a critical component in  program analysis for security, as understanding how information flows between the instructions of a program is often necessary to evaluate the program's security.
Unfortunately, static data-flow analysis is undecidable in the general case.
Moreover, when evaluating the security of a program, analyzing the source code is often considered to be insufficient due to the What You See is not What You Execute phenomenon~\cite{DBLP:journals/toplas/BalakrishnanR10}.
This creates the necessity for \textit{accurate data-flow analysis}  designed to operate \textit{on the binary instructions} of programs.

Fortunately, while data-flow analysis may be undecidable in a general case, there are many practical scenarios in which it is possible to perform data-flow analysis on small code segments (e.g., within a function), by implementing certain approximations in binary analysis engines~\cite{angr,ghidra,miasm}.
These approximations manifest as assumptions made in the model and heuristics incorporated into the implementation of the binary analysis engine. The approximations may at times sacrifice correctness in order to achieve guaranteed termination (and indeed scalability). Incorrect data-flow analysis manifests as false positives (non-existing data flows identified) and false negatives (existing data flows missed). Unfortunately, there is currently no systematic approach to evaluate the size and scope of inaccuracies in binary engines' implementations of data flow analysis, and to identify opportunities for improvements. 

This paper aims to benchmark data-flow approaches in binary analysis engines, to produce comprehensive, quantitative and insightful findings about their strengths and limitations. 
The benchmark dataset should not only measure accuracy of a data-flow analysis approach, but it should also identify root causes for inaccuracies (false positives and false negatives), which can be tackled by future researchers. We demonstrate how the insights from our evaluation led us to propose three improvements to angr's data-flow analysis model, which improved recall from 40\% to 99\% and precision from 13\% to 32\%.

Improving data-flow analysis accuracy is an essential step towards improving binary program analysis for system security. This paper demonstrates such an improvement in the context of vulnerability discovery, showing that our improvements to angr lead to higher identification of vulnerable data flows in three case studies.

In summary, this paper makes the following contributions.
\begin{itemize}
    \item We introduce \emph{alias classes}, a novel concept for categorizing data flows.
    \item We define and implement an open source framework for generating microbenchmark test cases, labeled with ground truth, that can be used to evaluate static data flow analysis models with respect to our alias classes.
    \item We define and implement a framework for extracting data flows from real-world programs using dynamic analysis, to generate ground-truth information about existing data flows.
    \item With the provided frameworks, we generate a data-flow evaluation benchmark data set, consisting of \nummbsrcs microbenchmark test cases mapping to \nummbfuncs unique binary executables, as well as data flows from $6$ real-world executables.
    \item We show the twofold utility of this data set:
        \begin{enumerate}
            \item We evaluate three state of the art static data-flow analysis  implementations in angr~\cite{angr}, Ghidra~\cite{ghidra} and Miasm~\cite{miasm}. Our evaluation yields insights about how accurate these analysis engines are with respect to different categories of data flows, and enables us to identify areas for improvement. To the best of our knowledge, this is the first evaluation of its kind.
            \item We propose three novel data-flow model extensions, implement them on top of angr, and evaluate them on real-world executables. These extensions achieve nearly perfect recall, while simultaneously improving precision of data-flow analysis on our benchmarks. 
            \item We show that leveraging our model extensions in a vulnerability-discovery context leads to an improved recovery of vulnerability-related instructions in three case studies.
        \end{enumerate}
\end{itemize} \mclearpage
\section{Binary-level data-flow analysis}
\label{sec:binary_data_flow_analysis}
To perform data-flow analysis on a program is to reason about the flow of information between its instructions.
Since data flow between two instructions requires executing these instructions in order, data-flow analysis is often built on top of control-flow analysis.
Consequently, data-flow analysis inherits the complexities of control-flow analysis, such as \textit{context sensitivity} -- the notion that the next instruction in an execution path may be determined by the preceding instructions.
For example, the instruction following a return instruction, is determined by the preceding, matching call instruction.
This complexity is part of a larger problem caused by indirect branches, in which case the control-flow depends on the data flow.
Due to the intermixed nature of control-flow and data-flow analysis, in the general case, both problems are undecidable~\cite{DBLP:journals/toplas/Reps00}. 
Moreover, performing these analyses on binary code, as opposed to source code, adds an extra layer of complexity due to the loss of high-level semantic information, such as control-flow structures, data types and data structures. However, binary code is the most faithful program representation when it comes to reasoning about a range of security properties, and it is therefore our target.

One of the root causes of the undecidability of data-flow analysis lies in the \textit{aliasing problem}, i.e., determining if two instructions access the same data (\textit{i.e.}, if two different pointers point to the same location).
In binary program analysis, this means determining if a memory write instruction and a subsequent memory read instruction access memory at the same address.
A data flow exists between these instructions if a control flow exists and if they must (or may) access the same data, such that the data written by the first instruction is read by the second instruction, and has not been changed in the meantime. 
Stated differently, a data flow exists between two instructions if they form a link in a def-use chain~\cite{DBLP:books/aw/AhoSU86}.

In spite of data-flow analysis being undecidable, a number of theoretical algorithms~\cite{DBLP:conf/scam/KissJLG03,DBLP:conf/vstte/BalakrishnanRMT05} and implementations of binary-level data-flow analysis~\cite{angr,ghidra,miasm} have been created to yield approximate solutions.
While the theoretical data-flow analysis algorithms make well-defined assumptions in order to guarantee termination, in the implementations it is necessary for developers to deviate from these theoretical models in order to achieve scalability in addition to termination.
This deviation manifests as additional, often undocumented, \textit{approximations} (assumptions and heuristics), which may lower analysis accuracy.
We discuss examples of such approximations in Section~\ref{sec:data_flow_analysis_implementations}.
In this paper we use our benchmarks to quantify the impact on accuracy from various approximations, which enables us to pinpoint areas for improvement, and to implement and demonstrate benefit of several improvements (Section~\ref{sec:improving_the_state_of_the_art}).

\subsection{Definitions}
In order to rigorously define our approach, we extend the existing concept of data flow with the following definitions.
\subsubsection{Degree of data flow}
We define three different degrees of data flow between any pair of instructions in a program: \emph{unconditional data flow}, \emph{possible data flow} and \emph{impossible data flow}.

A pair of instructions have an \textit{unconditional data flow}, if on every execution of the program, in which these instructions are executed in order, there is a data flow from the first instruction to the second.
We show an example of an unconditional data flow in Listing~\ref{list:univ_data_flow}.

\begin{center}
\noindent\begin{minipage}{0.90\columnwidth}
\begin{lstlisting}[language={[x64]Assembler}, caption={The instruction pair on lines~\ref{list:univ_data_flow:write} and~\ref{list:univ_data_flow:read} have an unconditional data flow.}, label={list:univ_data_flow}, escapechar=|]
mov [rdi], dl ; Write|\label{list:univ_data_flow:write}|
mov al, [rdi] ; Read|\label{list:univ_data_flow:read}|
\end{lstlisting}
\end{minipage}
\end{center}

A pair of instructions have a \textit{possible data flow}, if on at least one execution of the program there is a data flow from the first instruction to the second.
We show an example of a possible data flow in Listing~\ref{list:true_data_flow_1}.
In Listing~\ref{list:true_data_flow_1}, the data written to memory by the instruction on line~\ref{list:true_data_flow_1:write} will be read from memory by the instruction on line~\ref{list:true_data_flow_1:read} if and only if the value in register \texttt{rsi} is $0$.
If this value is dependent on input to the program, then in some executions there will be a data flow, while in others not.


\begin{center}
\noindent\begin{minipage}{0.90\columnwidth}
\begin{lstlisting}[language={[x64]Assembler}, caption={The instruction pair on lines~\ref{list:true_data_flow_1:write} and~\ref{list:true_data_flow_1:read} have a possible data flow.}, label={list:true_data_flow_1}, escapechar=|, boxpos=t]
mov [rdi], dl     ; Write|\label{list:true_data_flow_1:write}|
mov al, [rdi+rsi] ; Read|\label{list:true_data_flow_1:read}|
\end{lstlisting}
\end{minipage}\hfill
\end{center}

A pair of instructions have an \textit{impossible data flow}, if on every execution of the program, there is no data flow from the first to the second.
We show an example of an impossible data flow in Listing~\ref{list:false_data_flow}.
In Listing~\ref{list:false_data_flow}, the $1$ byte written to memory by the instruction on line~\ref{list:false_data_flow:write} will never be read from memory by the instruction on line~\ref{list:false_data_flow:read}.
Regardless of the value in the register \texttt{rdi} (used as the address in Line~\ref{list:false_data_flow:write}), this value can never be equal to $8$ less than itself (used as the address in Line~\ref{list:false_data_flow:read}).

\begin{center}
\noindent\begin{minipage}{0.90\columnwidth}
\begin{lstlisting}[language={[x64]Assembler}, caption={The instruction pair on lines~\ref{list:false_data_flow:write} and~\ref{list:false_data_flow:read} have an impossible data flow.}, label={list:false_data_flow}, escapechar=|]
mov [rdi], dl   ; Write|\label{list:false_data_flow:write}|
mov al, [rdi-8] ; Read|\label{list:false_data_flow:read}|
\end{lstlisting}
\end{minipage}
\end{center}

\subsubsection{Data flow scope}
We define the scope of a data flow as being either \emph{intra-procedural}, \emph{inter-procedural} or both.
A data flow is considered intra-procedural if it occurs in a single execution of a function.
%
%
Conversely, a data flow is inter-procedural if its instructions span multiple functions, or multiple executions of a single function.

\subsubsection{Data flow channel}
Given that a data flow is caused by a pair of instructions writing and reading the same data location, we define the \textit{channel of a data flow} as the register or memory that is accessed by both instructions.
We indicate this channel by using a token matching the name of the register or \texttt{mem} for memory.

We show an example of data flow channels in Listings~\ref{list:data_flow_chan_1} and~\ref{list:data_flow_chan_2}.
In Listing~\ref{list:data_flow_chan_1}, the instructions on lines~\ref{list:data_flow_chan_1:write} and~\ref{list:data_flow_chan_1:read} have the data flow channel \texttt{rbx} as they both access this register.
In Listing~\ref{list:data_flow_chan_2}, the instructions on lines~\ref{list:data_flow_chan_2:write} and~\ref{list:data_flow_chan_2:read} write and read memory at the same address respectively, so they have the data flow channel \texttt{mem}.

\begin{center}
\begin{minipage}{0.45\columnwidth}
\begin{lstlisting}[language={[x64]Assembler}, caption={The instruction pair on lines~\ref{list:data_flow_chan_1:write} and~\ref{list:data_flow_chan_1:read} have a data flow channel \texttt{rbx}.}, label={list:data_flow_chan_1}, escapechar=|]
mov rbx, 0   ; Write|\label{list:data_flow_chan_1:write}|
mov rax, rbx ; Read|\label{list:data_flow_chan_1:read}|
\end{lstlisting}
\end{minipage}\hfill
\noindent\begin{minipage}{0.45\columnwidth}
\begin{lstlisting}[language={[x64]Assembler}, caption={The instruction pair on lines~\ref{list:data_flow_chan_2:write} and~\ref{list:data_flow_chan_2:read} have a data flow channel \texttt{mem}.}, label={list:data_flow_chan_2}, escapechar=|]
mov [rbp-0x8], rdi ; Write|\label{list:data_flow_chan_2:write}|
mov rax, [rbp-0x8] ; Read|\label{list:data_flow_chan_2:read}|
\end{lstlisting}
\end{minipage}
\end{center}

\subsection{Our Scope}
\label{sec:scope}
In this paper, we separate the intermixed nature of control flow and data flow in order to focus exclusively on the latter.
We posit that the main challenge of data flow analysis, independently of control flow, is approximating a solution to the aliasing problem.
Therefore, in all our test cases, we focus on data flows between pairs of memory write and read instructions, i.e., with \texttt{mem} in the data-flow channel. 
We do not explicitly determine if a data-flow analysis is capable of identifying data flows between instructions writing and reading the same CPU register.
We argue that solving the aliasing problem in this particular case is trivial, since the location where the data is accessed (the register), is identifiable from the encoding of the instruction itself. 
We further focus on intra-procedural data-flow analysis, and leave inter-procedural data flows for future work.
We do this, because most data-flow analysis implementations limit their scope to intra-procedural analyses (in large part because of scalability issues caused by the limitations of existing pointer aliasing analysis models).


\subsection{Data-flow analysis implementations}
\label{sec:data_flow_analysis_implementations}

Due to the scale and complexity of modern software, developers of data-flow analysis implementations are required to introduce approximations in order to make the analysis usable on real-world executables.
This includes implementing heuristics to determine if a pair of instructions access the same memory, and if the data written by the first instruction is overwritten prior to being read by the second instruction.

A heuristic to determine if two instructions access the same memory must handle cases where at least one of the addresses is undefined in the scope of analysis.
We provide an example of this with Listing~\ref{list:true_data_flow_1}.
We illustrate the complexity of determining whether or not a memory value is overwritten with function calls.
Refer to Listings~\ref{list:callee_heuristic_1} and~\ref{list:callee_heuristic_2}.
In both these listings the \texttt{f\_target} functions are identical, writing to stack memory (labeled as \texttt{Write}) and subsequently reading from stack memory (labeled as \texttt{Read}).
However, these \texttt{Write} and \texttt{Read} instructions are interrupted by a function call to \texttt{f\_callee}.
In Listing~\ref{list:callee_heuristic_2} the callee function on line~\ref{list:callee_heuristic_2:callee} overwrites the same memory written by the \texttt{Write} instruction.
Therefore this listing shows an \textit{impossible data flow}.
On the other hand, in Listing~\ref{list:callee_heuristic_1} the callee function on line~\ref{list:callee_heuristic_1:callee} only reads this memory.
Therefore this listing shows an \textit{unconditional data flow}.
Since the effects of \texttt{f\_callee} are out of scope for an intra-procedural analysis of function \texttt{f\_target}, a data-flow analysis implementation must use a single heuristic to handle both these cases.
Note that any heuristic will therefore lead to inaccuracies for some \texttt{f\_target} implementations.
While in this case it would be trivial to include the instructions of \texttt{f\_callee} in the analysis (and perform inter-procedural data-flow analysis), in real-world programs functions tend to form large call graphs, with data-flows spanning multiple functions.
Any scalable data-flow analysis will necessarily need to limit its scope of analysis and use heuristics to reason about out-of-scope behavior.

In Sections~\ref{sec:evaluation} and~\ref{sec:improving_the_state_of_the_art} we investigate how data-flow analysis implementations approach these types of complexities and we propose alternative approaches that improve analysis accuracy.

\begin{minipage}{0.90\columnwidth}
\begin{lstlisting}[language={[x64]Assembler}, caption={The instructions on line~\ref{list:callee_heuristic_1:write} and line~\ref{list:callee_heuristic_1:read} have an unconditional data flow.}, label={list:callee_heuristic_1}, escapechar=|]
f_target:
  mov [rsp+0x8], 0   ; Write|\label{list:callee_heuristic_1:write}|
  lea rdi, [rsp+0x8]|\label{list:callee_heuristic_1:arg}|
  call f_callee
  mov rax, [rsp+0x8] ; Read|\label{list:callee_heuristic_1:read}|
  ret
f_callee:
  mov rax, [rdi]|\label{list:callee_heuristic_1:callee}|
  ret
\end{lstlisting}
\end{minipage}

\begin{minipage}{0.90\columnwidth}
\begin{lstlisting}[language={[x64]Assembler}, caption={The instructions on line~\ref{list:callee_heuristic_2:write} and line~\ref{list:callee_heuristic_2:read} have an impossible data flow.}, label={list:callee_heuristic_2}, escapechar=|]
f_target:
  mov [rsp+0x8], 0   ; Write|\label{list:callee_heuristic_2:write}|
  lea rdi, [rsp+0x8]|\label{list:callee_heuristic_2:arg}|
  call f_callee
  mov rax, [rsp+0x8] ; Read|\label{list:callee_heuristic_2:read}|
  ret
f_callee:
  mov [rdi], 0|\label{list:callee_heuristic_2:callee}|
  ret
\end{lstlisting}
\end{minipage} \mclearpage
\section{Designing a data set to benchmark static data-flow analysis}
Since data flow is undecidable in general~\cite{rice1953classes,DBLP:journals/toplas/Ramalingam94} the utility of any data-flow model can only be evaluated experimentally.
We propose a benchmark data set for evaluation of data-flow models, discussed in Section~\ref{sec:data_set}. 
We further implement an automated framework for evaluating data-flow models against our benchmarks, discussed in Section~\ref{sec:evaluation_overview}.

\subsection{Data Set}
\label{sec:data_set}
To gain a fine-grained insight into the efficacy of a data-flow analysis model, we break down data flows into a number of categories, which we refer to as \emph{alias classes}, discussed in Section~\ref{sec:alias_classes}.

Using these alias classes as a guide, we create a framework for generating a data set of test cases.
We divide these test cases into two categories: microbenchmark test cases, discussed in Section~\ref{sec:microbenchmark_test_cases}, and real-world test cases (Section~\ref{sec:real_world_test_cases}).
The microbenchmark test cases are synthesized at the source-code level and then compiled; the compilation process enables us to label them with ground truth data flows.
The real-world test cases are extracted from real-world programs.
We use a dynamic analysis approach to label them with ground truth about existing data flows (we mitigate side effects of the incompleteness of dynamic analysis coverage in  Section~\ref{sec:rw-testcases}).



After generation, this data set is a collection of binaries, each paired with information regarding its functions and data flows.
Each data flow is also assigned a degree of data flow as ground truth and an alias class.


\subsubsection{Alias classes}
\label{sec:alias_classes}
We categorize an intra-procedural data flow between a pair of instructions by how the pointers -- dereferenced by these instructions -- are introduced in the function at the source code level.\jelena{Nicolaas, can you say here in one sentence WHY we do this, like we do it so it can help us reason about possible changes to the pointers. Pointers of some types will usually not change by the program while others may easily change}
We refer to these categories as \emph{alias classes} and the introduction method of a pointer as the \emph{pointer origin}.
We define four such pointer origins, \emph{stack}, \emph{heap}, \emph{foreign} and \emph{global}.
For the \emph{foreign} pointer origin, the pointer is defined outside the function and introduced via a function argument.
For the \emph{stack} pointer origin, the pointer is introduced as an offset of the stack pointer register.
For the \emph{heap} pointer origin, the pointer is introduced via the return value of a memory allocation function.
Finally, \emph{global} pointers are allocated upon program initialization and are accessed either as a constant address, or an address relative to the instruction pointer.
We show a source-code level illustration of each of the pointer origins in Listing~\ref{list:pointer_origin_examples}.
Additionally, we show examples of data flows with alias classes \texttt{(Stack, Stack)} and \texttt{(Global, Foreign)} in Listings~\ref{list:alias_class_example_1} and~\ref{list:alias_class_example_2}, respectively.

\begin{minipage}{0.95\columnwidth}
\begin{lstlisting}[language={C}, caption={Pointers with each of the pointer origins.}, label={list:pointer_origin_examples}, escapechar=|]
char global_pointer;
void f(char *foreign_pointer) {
  char stack_pointer;
  char *heap_pointer = malloc(1);
}
\end{lstlisting}
\end{minipage}

\begin{minipage}{0.40\columnwidth}
\begin{lstlisting}[language={C}, caption={An unconditional data flow (line~\ref{list:alias_class_example_1:write} to~\ref{list:alias_class_example_1:read}) with alias class \texttt{(Stack, Stack)}.}, label={list:alias_class_example_1}, escapechar=|]
char f(char c) {
  char stack_ptr;
  stack_ptr = c; |\label{list:alias_class_example_1:write}|
  return stack_ptr; |\label{list:alias_class_example_1:read}|
}
\end{lstlisting}
\end{minipage}\hfill
\begin{minipage}{0.46\columnwidth}
\begin{lstlisting}[language={C}, caption={A possible data flow (line~\ref{list:alias_class_example_2:write} to~\ref{list:alias_class_example_2:read}) with alias class \texttt{(Global, Foreign)}.}, label={list:alias_class_example_2}, escapechar=|]
char global_ptr;
char f(char *foreign_ptr) {
  global_ptr = c; |\label{list:alias_class_example_2:write}|
  return *foreign_ptr; |\label{list:alias_class_example_2:read}|
}
\end{lstlisting}
\end{minipage}

\jelena{please use pictures in this WHOLE section. Start with simple memory write followed by read, then draw how you expand this by alias classes, by pointer types and sizes, by compiler options, etc. Draw from left to right, label each expansion with a letter like (a), (b). Then refer to this picture throughout this section, like "Figure 3(b) shows".}

\subsubsection{Microbenchmark test cases}
\label{sec:microbenchmark_test_cases}
The purpose of our microbenchmark test cases is to span a wide variety of intra-procedural data flows that can occur in programs, thus enabling comprehensive evaluation of data-flow analysis. 
Each test case is designed to be minimalistic, testing a single target data flow in a target function, focused solely on executing this data flow.

At the heart of each test case lie a pair of instructions writing and reading memory addresses, forming the target data flow.
The ground truth of each test case -- whether there exists an unconditional, possible or impossible data flow between these instructions -- depends on the parameters of its construction.
The source code of the test cases are generated by enumerating these parameters.
Finally, each instance of the source code is compiled in multiple ways, enumerating compiler options, and producing one binary per option set.

\paragraph{Pointer creation}
The first phase of test case creation is to create the pointers that will be used by the memory access instructions.
This phase involves \emph{pointer definition} and \emph{pointer expansion}.
In creating the data set, we enumerate a set of properties that make up the pointer definition -- a pointer origin, data type, size and length -- and expand the pointer into a write and read pointer.
The data type is a native data type supported by the compiler and underlying architecture, e.g. integer or floating point.
The size attribute of a pointer indicates the number of bits comprising the data type.
Finally, the length describes the number of adjacent data types in memory the pointer points to.
For a length greater than $1$, the pointer points to an array.
These properties of a pointer are used to define it in the source code of the test case, including allocating the necessary amount of space on the stack or heap.

In the next phase, \emph{pointer expansion}, we either use a single pointer for both the write and read instruction, or two distinct pointers for each instruction.
We refer to the pointer used in the write and read instruction as the \emph{write pointer} and \emph{read pointer}, respectively, regardless of whether or not they are the same.
We use a single pointer for both the write and read pointer in order to construct unconditional data flows.

\paragraph{Pointer transformation}
\label{sec:pointer_transformation}
To increase test case complexity and variety, we introduce an optional pointer transformation.
This transformation adds an offset to the pointer, which can either be a constant value, or a variable undefined within the target function.
The length attribute of the pointer is used to select an offset within the bounds of the pointer array.
The transformed pointers are ultimately used in the target data flow. \jelena{another good place to insert a figure, either use the one I suggested above or insert new one to illustrate}

\paragraph{Callee interruption}
As discussed in Section~\ref{sec:data_flow_analysis_implementations}, an intra-procedural data-flow analysis must use approximations in order to handle function calls in the target function.
In order to expose these approximations, we create a counterpart for each test case where the target data flow is interrupted by a function call.
We show an example of such a pair of test cases in Section~\ref{sec:improving_the_state_of_the_art} in Listings~\ref{list:unconditional_no_callee} and~\ref{list:underspecified_callee}. 

\paragraph{Test case compilation}
We increase test case diversity by compiling the source code with a variety of compiler flags.
This is useful, because compiler flags such as optimization options have a significant impact on how the source code is converted into binary code.
Additionally, options to include or omit the frame pointer have an effect on how stack variables are represented in binary code. Different binary code representations can may impact accuracy of data flow analysis in different binary analysis engines.

\paragraph{Ground truth}
\label{sec:microbenchmark_test_cases_ground_truth}
With respect to ground truth, we divide the microbenchmark test cases into two subcategories, the \textit{fully-specified} test cases and the \textit{underspecified} test cases. 

In \textit{fully-specified test cases} all information regarding the existence of the target data flow is within the target function. 
In every fully-specified test case, the target data flow either occurs on every execution of the program (unconditional data flow) or does not occurs on any execution of the program (impossible data flow).
Therefore, any result reported by a data-flow analysis that contradicts the ground truth can be confirmed as an error. The goal of fully-specified test cases is to uncover the concrete strengths and weaknesses of a particular approach.
Listings~\ref{list:univ_data_flow} and~\ref{list:false_data_flow} are examples of fully-specified test cases.

\textit{Underspecified test cases} include a target data flow that may or may not occur, depending on data outside of the target function.
Listing~\ref{list:true_data_flow_1} shows an example of an underspecified test case.
Underspecified test cases have \emph{possible} data flows, because by definition, the information missing from the scope of analysis can be defined specifically to cause a data flow.
In the example shown in Listing~\ref{list:true_data_flow_1}, a data flow exists when register \texttt{rsi} is equal to $0$.
Note, however, that in some cases a possible data flow is unlikely.
For example, between an instruction writing to global memory and an instruction reading from stack memory.
For a data flow to exist, the stack pointer would need to point to global memory.
The purpose of the underspecified test cases, is to evaluate how data-flow analysis works in scenarios where perfect information is unavailable. Many real-world binaries contain instances of functions with underspecified data flows. 

\subsubsection{Real-world Test Cases}
\label{sec:real_world_test_cases}
Our microbenchmark test cases are intentionally  simple and unambiguous in order to reflect important program properties. However, these are not meant to reflect the complexity of real-world programs. 
For this reason, we also enrich our benchmarks with test cases built from real-world binaries. To accomplish this we must address three challenges.
Firstly, we need to establish ground truth, i.e. which data flows exist within a given real-world binary.
Secondly, we need to construct intra-procedural data-flow graphs composed of these ground-truth data flows.
Thirdly, we need to determine the alias class of each of these data flows.

\paragraph{Establishing ground truth}
Unlike the microbenchmark test cases, where we construct the ground truth in the source code and compile, in real-world test cases the ground truth must be inferred.
Dynamic analysis allows us to establish that a data flow exists in some executions of the binary.
Therefore, dynamic analysis is capable of uncovering possible data flows in a binary, but it cannot identify unconditional or impossible data flows.
By definition, dynamic analysis considers a target program one program path at a time.
Since real-world program are large, they potentially have an unbounded number of program paths, making  dynamic analysis incomplete.
Consequently, our ground truth includes \textit{a number of} possible data flows, which may be incomplete. We address this in Section~\ref{sec:rw-testcases} by computing lower and upper bound approximations for our real-world test cases. 

In order to establish ground truth for a target binary, we perform dynamic instrumentation and log all information necessary to recover the data flows.
This information includes the instruction address, data access type (write or read), the data location accessed (either a register identifier, or memory address) and the context (the function call in which the instruction executes).
Whenever a data write access is encountered, a map is used to associate the data location with the instruction that performed the access.
Whenever a data read access occurs, this map is consulted to identify the matching write access.
The result is a pair of instructions, for which the first writes to the same data location that is read by the second, thus forming a data flow.

\paragraph{Creating the data-flow graphs}
The dynamically-collected data flows are consolidated into an inter-procedural data-flow graph, where instruction addresses are the nodes and directed edges indicate a data flow.
We annotate the edges in this graph with information pertaining to the scope and channel of the data flow.

In order to separate the inter-procedural data-flow graph into an intra-procedural subgraph we extract a node-induced subgraph, using the instructions of the target functions as the node set.
We also eliminate all edges in this subgraph that correspond to a data flow with only an inter-procedural scope.

\begin{figure}
    \centering
    \includegraphics[width=1\columnwidth, keepaspectratio]{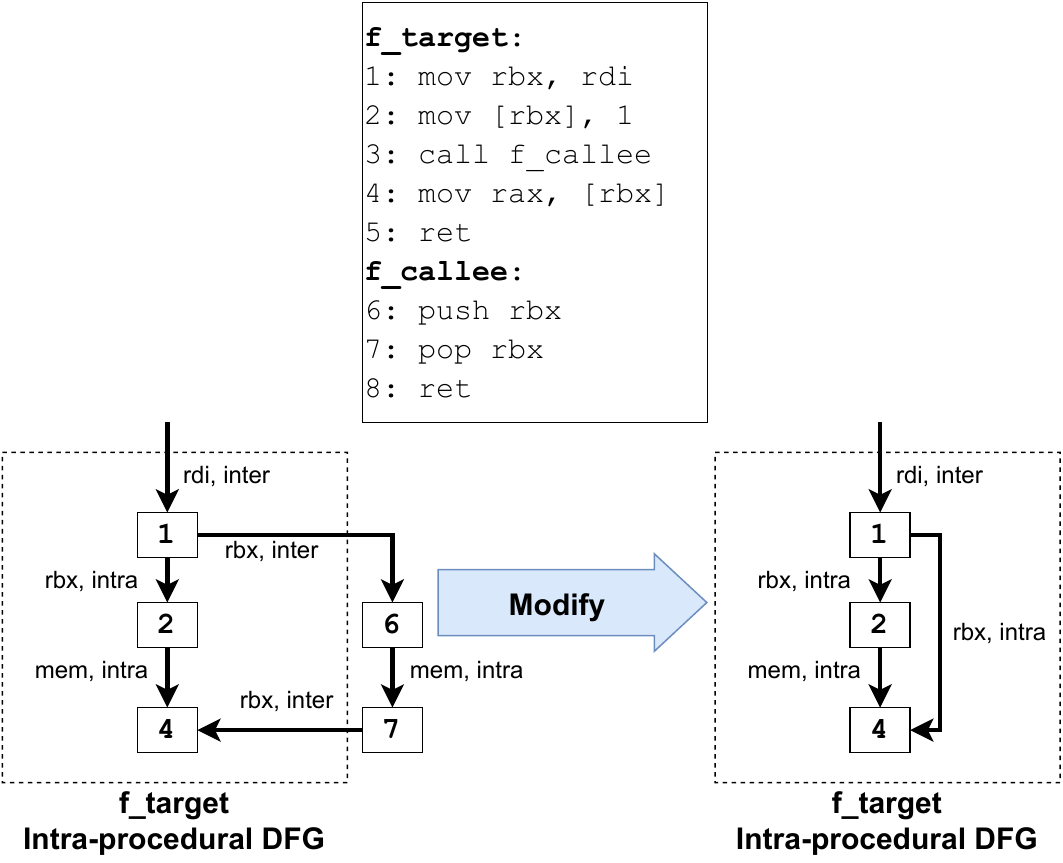}
    \caption{We modify the intra-procedural data-flow graph to reconnect registers across save-restore edges.}
     \label{fig:dynamic_dfg_mod_save_restore}
     
\end{figure}

\paragraph{Handling special cases}
We apply the following modifications to the dynamically-generated intra-procedural data-flow graph to handle common special cases found in binary programs.
A common pattern in binary code is to preserve registers across function calls, by writing the register to memory at the start of the function and then reading it back again into the register at the end.
In our dynamically-generated intra-procedural data-flow graphs, this appears as undefined registers used after the function call. 
We identify these save-restore patterns in the inter-procedural data-flow graph and reconnect the instruction addresses in the target function with an intra-procedural data-flow edge.
We show an example of this in Figure~\ref{fig:dynamic_dfg_mod_save_restore}.
The register \texttt{rbx} is saved across the function call \texttt{f\textunderscore callee} and therefore, we reconnect the definition and use of this register in \texttt{f\_target}.

\begin{figure}
    \centering
    \includegraphics[scale=0.4, keepaspectratio]{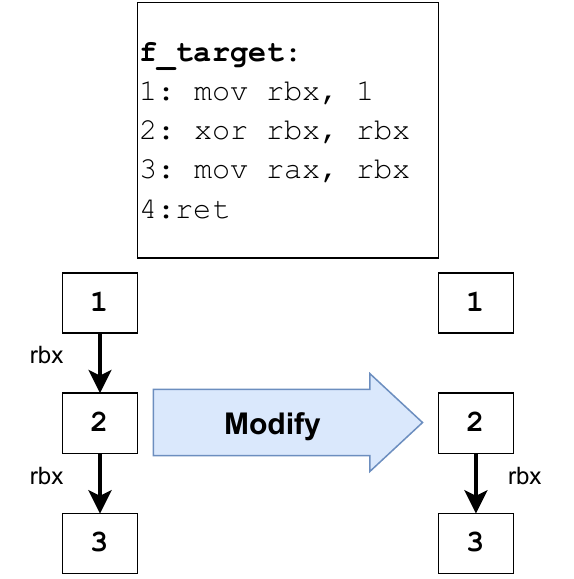}
    \caption{We modify the intra-procedural data-flow graph to eliminate data flows into register clearing instructions.}
     \label{fig:dynamic_dfg_mod_xor}
     
\end{figure}

The second modification we make, is to identify and remove data flows leading into an instruction used to clear a register. 
We show an example of this in Figure~\ref{fig:dynamic_dfg_mod_xor}.
The \texttt{xor} instruction clears the value in the register \texttt{rbx}.
Therefore, even though this instruction writes and reads this register, no data flow exists through this instruction.
If such an instruction is used to clear a register at the start of the function, in the intra-procedural data-flow graph, any subsequent instruction that reads this register will appear to use an undefined register. 
Instead, the register-clearing instruction should be interpreted as setting the register to $0$, with no incoming data flows. 

We make these modifications, because correctly identifying registers undefined in the target function forms the basis of how we identify alias classes in real-world binaries.
We explain this next. 

\paragraph{Alias class identification}
Recall that the alias class of a data flow is defined by the origin of the pointers dereferenced in data write and read instructions.

Identifying pointer origins in real-world program is challenging, because between its introduction and use, the pointer may be assigned to a different register and also saved (restored) to (from) memory.
Therefore, a simple syntactic analysis of the pointer is insufficient to correctly determine its origin.
Indeed, it is necessary to trace the data flow of a pointer back from its use to its introduction into the function.
We identify the introduction point of an address by following its intra-procedural data flow backwards until an inter-procedural data flow is encountered.
The undefined registers upon which this address depends is the union of the data-flow channels of these inter-procedural edges. 

For a stack pointer origin, the pointer is introduced via the stack pointer register, which is undefined within a function (as its value depends on the call stack).
Foreign pointers are passed to the target function as arguments.
How this is implemented in binary code, depends on the calling convention.
In this paper, we focus on the calling convention specified in the System V AMD64 Binary Application Interface.
In this calling convention, pointer arguments are passed in the registers \texttt{rdi}, \texttt{rsi}, \texttt{rdx}, \texttt{rcs}, \texttt{r8}, \texttt{r9} (additional pointer arguments are passed via the stack).
Therefore, a foreign pointer is introduced via one of these undefined registers.
Heap pointers are most often obtained as the return value of a memory allocation function (e.g., \texttt{malloc}), which is passed via the \texttt{rax} register.
Therefore, heap pointers are introduced via an undefined \texttt{rax} register.
In the case of global pointers, the address does not depend on any undefined registers, because the address is constant or it is an offset of the instruction pointer.

\subsection{Evaluation Framework}
\label{sec:evaluation_overview}
We evaluate a data-flow analysis approach by applying it on the binaries, functions and data flows in our benchmark data set.
We refer to the binary and function under evaluation as the \emph{target binary} and \emph{target function}, respectively.
Similarly, the data flow under test is is the \emph{target data flow}.
For each target function, we generate a data-flow graph using the binary analysis engine (e.g., angr) under evaluation.
This data-flow graph is examined to determine the presence (or absence) of the edge between the two instructions constituting the target data flow.
We compare this information  to the ground truth to establish correctness. \mclearpage
\section{Implementation}
\label{sec:implementation}

\subsection{Selected approaches}
We evaluate the three state-of-the-art data-flow analysis implementations found in the binary program analysis engines angr~\cite{angr} (version 9.2.39), Ghidra~\cite{ghidra} (version 10.2.2) and Miasm~\cite{miasm} (version 0.1.3.dev447).
We refer to these as our \emph{selected approaches}.
These binary analysis engines are popular in both academia and the industry, and were selected for being open source, and for providing an  implementation of a general-purpose, best-effort data-flow analysis.
In Section~\ref{sec:evaluation}, we conduct our evaluation using these selected approaches to uncover the similarities and differences between these approaches, as well as their strengths and weaknesses.

\subsection{Microbenchmark test cases}
We implement an open source framework\footnote{Anonymized for review} to generate our microbenchmark test cases automatically by following the approach in Section~\ref{sec:microbenchmark_test_cases}. \jelena{here you would either summarize these option in a table and refer to it, or you would use the figure I suggested above and refer to it. I think you need both a table here and a figure earlier.}
For the pointer definition properties, we use the native data types and sizes: 8-bit \texttt{int}, 16-bit \texttt{int}, \texttt{float} and \texttt{double}.
We also include a defined \texttt{struct} data type comprising an integer and pointer value.
For the length property of pointers, we assign one of two constant values, $1$ and $2$, or one of two variable values.
These variables are passed to the target function via function parameter.
We incorporate these variable lengths to test undefined offset transformations.

After generating the source code, we generate the test case binaries by compiling the source code with the GNU Compiler Collection (GCC, version 11.3.0) using one of the $6$ optimization options (\texttt{O0-O3}, \texttt{Os} and \texttt{Ofast}) as well varying the inclusion of the function stack frame pointer with \texttt{-fomit-frame-pointer}. 
All pointers are tagged as \texttt{volatile} in the source code to prevent the compiler optimizing away the target data flow.
Every target binary is compiled with DWARF debug symbols, using the flag \texttt{-gdwarf-4}.
The debug symbols are used to identify the memory access instructions of the target data flow in the compiled binary, using the \texttt{pyelftools} Python library~\cite{pyelftools}, and to establish ground truth for data flows in our microbenchmarks.
Finally, given the set of compiled binaries, a fingerprint is calculated for each target function by computing the MD5-hash of the bytes comprising its machine code instructions.
The MD5-hash is used to identify and eliminate duplicate functions.
The final set of microbenchmark test cases is a set of binaries, each containing a unique target function.
In total, we have \nummbsrcs source code level test cases, mapping to \nummbfuncs unique target functions.

\subsection{Real-world test cases}
For the real-world test cases, we select binaries from the following projects: \texttt{chmod}, \texttt{cp}, \texttt{ls} (from the Coreutils package~\cite{coreutils}, version v9.1-98-g8613d35be) as well as the \texttt{Apatche-Httpd} server~\cite{httpd} (version 2.5.1-dev), the \texttt{Mujs} javascript interpreter~\cite{mujs} (version  1.3.3) and \texttt{CJson} parser~\cite{cjson} (version  v1.7.15-14-gb45f48e).
We select these binaries to cover a range of types of application domains and complexities of software.

To dynamically collect execution traces for a target binary, we instrument it using Intel's PIN framework~\cite{DBLP:conf/pldi/LukCMPKLWRH05}, while executing test cases in the project repository.
Since the purpose of test cases is to cover a broad range of program functionality, these will allow us to recover data flows across many execution paths. 
We normalize instruction addresses, to compensate for address space layout randomization (ASLR).

From each of the real-world binaries, we select a subset of target functions, for which we will compute static data-flow graphs in our evaluation.
Specifically, we select the $5$ functions for which our dynamic analysis approach identified the highest number of memory data flows.
We show these functions for each target binary in Table~\ref{tab:real_world_alias_classes}, with the number of target data flows per alias class.
We denote the alias class as \emph{unknown} in the cases where our alias class identification for real-world binaries fails (see Section~\ref{sec:real_world_test_cases}).
We use these memory data flows as ground truth for the static data-flow analyses.

While we could perform our evaluation on all functions in the target binary, this is often a computationally expensive procedure.
We wanted to allocate sufficient time for selected static-analysis approaches to compute the data-flow graph, while still keeping the overall run time manageable.
To this end, we select the above-mentioned $5$ functions from each target binary and allow a $5$ hour time limit per target function. 

\begin{table}

\begin{center}
\scriptsize
\caption{The selected target functions for each real-world binary with the number of identified data flows per alias class.}
\label{tab:real_world_alias_classes}
\scalebox{0.90}{
\begin{tabular}{ l r r r r r }
 Target Function & (F, F) & (G, G) & (H, H) & (S, S) & Unknown  \\
 \hline
 \multicolumn{6}{c}{\texttt{chmod}}\\\hline
 \texttt{main} & 0 & 16 & 1 & 29 & 0 \\
 \texttt{quotearg\_buffer\_restyled} & 0 & 0 & 0 & 37 & 0 \\
 \texttt{fts\_build} & 0 & 0 & 0 & 23 & 3 \\
 \texttt{rpl\_fts\_open} & 0 & 0 & 1 & 4 & 0 \\
 \texttt{quotearg\_n\_options} & 0 & 0 & 0 & 4 & 0\\
 \hline
 \multicolumn{6}{c}{\texttt{cp}}\\\hline
 \texttt{copy\_internal} & 0 & 2 & 0 & 354 & 4 \\
 \texttt{sparse\_copy} & 1 & 0 & 0 & 44 & 5 \\
 \texttt{main} & 0 & 1 & 0 & 47 & 0 \\
 \texttt{backupfile\_internal} & 0 & 0 & 0 & 39 & 0 \\
 \texttt{make\_dir\_parents\_private} & 3 & 0 & 2 & 27 & 2 \\
 \hline
 \multicolumn{6}{c}{\texttt{ls}}\\\hline
 \texttt{main} & 0 & 66 & 0 & 57 & 0 \\
 \texttt{quotearg\_buffer\_restyled} & 0 & 0 & 0 & 52 & 0 \\
 \texttt{mpsort\_with\_tmp.part.0} & 1 & 0 & 0 & 37 & 8 \\
 \texttt{canonicalize\_filename\_mode} & 0 & 0 & 1 & 34 & 2 \\
 \texttt{\_\_strftime\_internal.isra.0} & 0 & 0 & 0 & 28 & 0\\
 \hline
 \multicolumn{6}{c}{\texttt{Apache-Httpd}}\\\hline
 \texttt{trie\_node\_link} & 9 & 0 & 0 & 53 & 0 \\
 \texttt{ap\_add\_module} & 0 & 0 & 0 & 39 & 2 \\
 \texttt{trie\_node\_alloc} & 0 & 0 & 1 & 36 & 0 \\
 \texttt{register\_filter} & 1 & 0 & 0 & 36 & 0 \\
 \texttt{ap\_setup\_prelinked\_modules} & 0 & 11 & 0 & 24 & 0\\
 \hline
 \multicolumn{6}{c}{\texttt{Mujs}}\\\hline
 \texttt{jsR\_run} & 4 & 0 & 0 & 380 & 6 \\
 \texttt{cstm} & 0 & 0 & 0 & 323 & 0 \\
 \texttt{jsC\_cexp} & 0 & 0 & 0 & 256 & 0 \\
 \texttt{js\_gc} & 2 & 0 & 0 & 197 & 0 \\
 \texttt{statement} & 2 & 0 & 0 & 177 & 0 \\
\hline
 \multicolumn{6}{c}{\texttt{CJson}}\\\hline
 \texttt{cJSON\_Delete} & 0 & 0 & 0 & 23 & 0 \\
 \texttt{get\_object\_item} & 0 & 0 & 0 & 18 & 0 \\
 \texttt{add\_item\_to\_object} & 0 & 0 & 0 & 18 & 0 \\
 \texttt{add\_item\_to\_array} & 0 & 0 & 0 & 16 & 0 \\
 \texttt{UnityAssertEqualString} & 0 & 0 & 0 & 15 & 0
\end{tabular}
}
\end{center}

\end{table}

\subsection{Evaluation framework implementation}
We implement the evaluation framework in an open source repository\footnote{Anonymized for review}, by creating a Python wrapper script for each selected approach (angr, Miasm, Ghidra) to generate a data-flow graph for a specified target function in a target binary.
Each of the selected approaches provide a number of settings to fine-tune the data-flow analysis for a particular use case. 
We make a best-effort approach to apply the settings that will maximize performance on our data set.
This is achieved by performing an analysis with a variety of settings, and selecting the best results per binary analysis engine to report. 
Next, we discuss these settings as well as the process of converting the approach-specific data-flow information into an unified-representation comparable to ground truth.

\paragraph{angr.}
By default, angr builds an inter-procedural data-flow graph on top of a control-flow graph (CFG), augmented with symbolic execution.
In order to focus this data-flow analysis on the target function, we reduce the scope of CFG generation to only this function.
We achieve this by disabling context sensitivity and setting the call-depth parameter to $0$.
angr produces a data-flow graph over the statements of the Vex IR.
Each of the statements in this IR is associated with a machine code instruction via an \texttt{IMark}\footnote{An \texttt{IMark} statement is a special statement in Vex IR that does not describe the behavior an instruction, but instead its address and size in bytes.} statement, which contains the instruction address of this machine code instruction.
We use these \texttt{IMark} statements to convert the data-flow graph produced by angr, to one over machine code instructions.

\paragraph{Ghidra.}
Ghidra performs its data-flow analysis in the process of converting the machine code instructions to its P-code IR.
In this process, the inputs and output of P-code operations are linked.
Each of these P-code operations are associated with a machine code instruction address, which we use to create a data-flow graph.
Per its default settings, Ghidra performs a number of simplification steps on the produced P-code, such as consolidating some operations.
This simplification eliminates the mapping between some machine code instructions and their corresponding P-code operations.
To prevent this, we set the simplification style to \texttt{firstpass} instead.

\paragraph{Miasm.}
Similar to angr, for Miasm we first create a CFG for the target function over its intermediate representation.
Then, we generate a dependency graph, while instructing Miasm explicitly to consider memory dependencies, while disregarding function calls. \mclearpage
\section{Evaluation}
\label{sec:evaluation}
We evaluate the selected approaches on both the microbenchmark test cases and real-world test cases of our data set.

\subsection{Microbenchmark test cases}
Our microbenchmarks consist of \nummbfuncs target binaries, each paired with information regarding its target function, target data flow and ground truth.
We use our evaluation framework to extract a static data-flow graph for the target function, using each of the selected approaches.
We inspect the produced data-flow graphs to determine if the target data flow is reported or not by each approach.
We compare this report with the ground truth for the test case.
We show the performance of the selected approaches on our microbenchmark test cases in Table~\ref{tab:microbenchmark_evaluation}.
Here, we only show the results of the fully-specified test cases because these have a clear ground truth.
In Section~\ref{sec:improving_the_state_of_the_art}, we show angr's performance on some underspecified test cases and we show how we leverage this to improve performance. 

In Table~\ref{tab:microbenchmark_evaluation} we observe that angr is capable of identifying every unconditional data flow in each alias class.
However, it also reported a data flow for each of the impossible data flows in the \texttt{(Foreign, Foreign)} and \texttt{(Heap, Heap)} alias classes and some ($20.19\%$) of the impossible data flows in the \texttt{(Stack, Stack)} alias class.
As all these data flows are impossible, each of these cases represents a false positive.
In Section~\ref{sec:improving_the_state_of_the_art} we conduct an investigation on these false positives and show how they indicate an opportunity for improvement.
We also investigate the few cases in which Ghidra reports a data flow.
We conclude that Ghidra does not perform alias analysis, it assumes all memory addresses are unequal.
There are a few exceptions in which it equates memory addresses in global memory, but we leave the deep-dive into Ghidra's source code to establish the reason for this as future work.
Finally, we observe that Miasm sporadically reports unconditional and impossible data flows.
Miasm performs alias analysis by syntactically comparing the memory access instructions.
This is a reasonable heuristic, but it is also sensitive to how the compiler implements the access instructions, especially with respect to register allocation.

\begin{table} 
\begin{center}
\scriptsize
\caption{Performance of selected approaches on the microbenchmark test cases}
\label{tab:microbenchmark_evaluation}
\scalebox{0.90}{
\begin{tabular}{ l l r r r r r}
 Alias Class & Ground Truth & Edge & Edge \% & No Edge & No Edge \% & Total  \\
 \hline
 \multicolumn{7}{c}{angr} \\\hline
 (F, F) & unconditional & $158$  & $100.00\%$  & $0$  & $0.00\%$  & $158$\\
 (F, F) & impossible & $72$ & $100.00\%$  & $0$  & $0.00\%$  & $72$ \\
 (G, G)   & unconditional & $170$ & $100.00\%$ & $0$  & $0.00\%$  & $170$\\
 (G, G)   & impossible     & $0$  & $0.00\%$  & $3,726$ & $100.00\%$  & $3,726$\\
 (H, H) & unconditional & $376$ & $100.00\%$ & $0$  & $0.00\%$  & $376$\\
 (H, H) & impossible     & $12,988$ & $100.00\%$ &  $0$ & $0.00\%$  & $12,988$\\
 (S, S) & unconditional & $115$ & $100.00\%$ & $0$  & $0.00\%$  & $115$\\
 (S, S) & impossible     & $717$ & $20.19\%$ & $2,835$ & $79.81\%$ & $3,552$\\
\hline\multicolumn{7}{c}{Ghidra} \\\hline
(F, F) & unconditional & $0$  & $0.00\%$ & $158$  & $100.00\%$  & $158$\\
 (F, F) & impossible & $0$  & $0.00\%$ & $72$ & $100.00\%$    & $72$ \\
 (G, G)   & unconditional & $10$ & $5.88\%$ & $160$  & $94.12\%$  & $170$ \\
 (G, G)   & impossible     & $210$  & $5.64\%$  & $3,516$ & $94.36\%$  & $3,726$ \\
 (H, H) & unconditional & $0$  & $0.00\%$ & $376$ & $100.00\%$ &   $376$\\
 (H, H) & impossible     &  $0$ & $0.00\%$ & $12,988$ & $100.00\%$   & $12,988$\\
 (S, S) & unconditional & $0$  & $0.00\%$ & $115$ & $100.00\%$   & $115$\\
 (S, S) & impossible     & $0$ & $0.00\%$ & $3,552$ & $100.00\%$ & $3,552$\\
\hline\multicolumn{7}{c}{Miasm} \\\hline
 (F, F) & unconditional & $106$ & $67.09\%$   & $52$  & $32.91\%$  & $158$\\
(F, F) & impossible     & $24$ & $33.33\%$   & $48$  &  $66.67\%$  & $72$\\
(G, G)   & unconditional & $52$ & $30.59\%$   & $118$ &  $69.41\%$   & $170$\\
(G, G)   & impossible     & $232$ &  $6.23\%$  &   $3,494$ & $93.77\%$ & $3,726$\\
(H, H) & unconditional & $376$ & $100.00\%$ & $0$  & $0.00\%$ & $376$\\
(H, H) & impossible     & $2462$ & $18.96\%$   &    $10,526$ & $81.04\%$ & $12,988$\\
(S, S) & unconditional & $107$ & $93.04\%$   &  $8$ & $6.96\%$    & $115$\\
(S, S) & impossible     & $150$ &  $4.22\%$  &  $3,402$  & $95.78\%$ & $3,552$
 \end{tabular}
}
\end{center}

\end{table}

\subsection{Real-world test cases}
\label{sec:rw-testcases}
Our real-world test cases consist of $6$ real-world binaries, each paired with information regarding $5$ target functions and dynamically recovered data flows.
For each target function, we compute the static data-flow graph using each of the selected approaches.
We show the run time of this process in Table~\ref{tab:real_world_static_dfg_times}.
Miasm fails to produce a data-flow graph for 6 target functions, due to exceeding a memory limit of $100$GB.
By using a profiler, we establish that the excessive memory usage is related to its task list of states to process.
These states are differentiated by code location and data flows present at that state.
Consequently, the size of this task list may grow disproportionately larger than the size of the target function.

\begin{table} %
\begin{center}
\scriptsize
\caption{Extraction time of the data-flow graph for the target function in each real-world binary.}
\label{tab:real_world_static_dfg_times}
\begin{tabular}{ l r r r }
 Target Function & angr & Ghidra  & Miasm  \\
 \hline
 \multicolumn{4}{c}{chmod} \\\hline
 \texttt{main}   & $14.63$s & $12.59s$ & OOM \\
 \texttt{quotearg\_buffer\_restyled} & $60.79$s & $13.08s$ & OOM \\
 \texttt{fts\_build} & $16.29$s & $12.21s$ & OOM \\
 \texttt{rpl\_fts\_open} & $10.60$s & $12.11s$ & $580.53$ \\
 \texttt{quotearg\_n\_options} & $9.38$s & $11.90s$ & OOM \\
 \hline
 \multicolumn{4}{c}{cp} \\\hline
 \texttt{copy\_internal} & $87.95$s & $16.32s$ & OOM \\
 \texttt{sparse\_copy} & $17.36$s & $14.62s$ & OOM \\
 \texttt{main} & $16.67$s & $14.92s$ & $54.52s$ \\
 \texttt{backupfile\_internal} & $15.90$s & $14.60s$ & OOM \\
 \texttt{make\_dir\_parents\_private} & $16.26$s & $14.27s$ & OOM \\
 \hline
 \multicolumn{4}{c}{ls} \\\hline
 \texttt{main} & $30.81$s & $18.98s$ & OOM \\
 \texttt{quotearg\_buffer\_restyled} & $66.94$s & $17.99s$ & OOM \\
 \texttt{mpsort\_with\_tmp.part.0} & $17.37$s & $16.75s$ & $249.54s$ \\
 \texttt{canonicalize\_filename\_mode}& $18.85$s & $17.03s$ & OOM \\
 \texttt{\_\_strftime\_internal.isra.0} & $77.55$s & $17.53s$ & OOM \\
 \hline
 \multicolumn{4}{c}{Apache-Httpd} \\\hline
 \texttt{trie\_node\_link} & $112.62$s & $43.09s$ & $2.40s$ \\
 \texttt{ap\_add\_module} & $112.51$s & $44.51s$ & $6.91s$ \\
 \texttt{trie\_node\_alloc} & $111.97$s & $43.42s$ & $1.11s$ \\
 \texttt{register\_filter} & $112.21$s & $43.97s$ & $1.65s$ \\
 \texttt{ap\_setup\_prelinked\_modules} & $112.47$s & $43.51s$ & $1.66s$ \\
 \hline
 \multicolumn{4}{c}{Mujs} \\\hline
 \texttt{jsR\_run} & $50.05$s & $22.39s$ & $0.73s$ \\
 \texttt{cstm} & $18.90$s & $21.94s$ & $0.48s$ \\
 \texttt{jsC\_cexp} & $24.92$s & $21.17s$ & $0.44s$ \\
 \texttt{js\_gc} & $17.45$s & $21.03s$ & OOM \\
 \texttt{statement} & $21.69$s & $21.70s$ & $19.81s$ \\
 \hline
 \multicolumn{4}{c}{CJson} \\\hline
 \texttt{cJSON\_Delete} & $5.05$s & $10.63s$ & $0.78s$ \\
 \texttt{get\_object\_item} & $4.72$s & $10.81s$ & $0.63s$ \\
 \texttt{add\_item\_to\_object} & $4.85$s & $10.80s$ & $0.55s$ \\
 \texttt{add\_item\_to\_array} & $4.64$s & $11.01s$ & $0.41s$ \\
 \texttt{UnityAssertEqualString} & $4.91$s & $10.56s$ & $0.62s$
\end{tabular}
\end{center}
\end{table}

We combine the data flows of each target function in a particular binary and show the total number per alias class in Table~\ref{tab:real_world_performance}.
Additionally, this table shows how many of the dynamic data flows are discovered by the selected approaches and the number of data flows reported by static analysis only.
We see that Ghidra surprisingly identifies data flows in more alias-classes than during the microbenchmark evaluation (Table~\ref{tab:microbenchmark_evaluation}).
However, after manual analysis, we conclude that these data flows exist between synthetic P-code operations inserted by Ghidra and do not reflect the target data flow.

We use the numbers in Table~\ref{tab:real_world_performance} to compute an aggregated performance score for each selected approach on real-world binaries. 
For each selected approach $\alpha$, we consolidate all data flows discovered dynamically and statically into a sets $D$ and $S_{\alpha}$, respectively.
We consider the intersection ($D \cap S_{\alpha}$) the true positive data flows discovered by $\alpha$.
As discussed in Section~\ref{sec:real_world_test_cases}, dynamic analysis is incomplete.
Every data flow reported only by static analysis ($S_{\alpha}\setminus D$ ) may either be a false positive, or a true positive for which we do not have dynamic evidence.
We assume each such data flow is a false positive.
Consequently, the number of true positives $|D \cap S_{\alpha}|$ and false positives $|S_{\alpha}\setminus D|$ we report is a lower and upper bound approximation, respectively.
To estimate the number of false negatives, we identify the dynamic data flows not discovered statically ($D \setminus S_{\alpha}$).
This is a lower-bound approximation, because there may be data flows unreported by both dynamic and static analysis.
We show these approximations in Table~\ref{tab:real_world_totals} along with an approximation of the precision, recall an \fscore score.
Table~\ref{tab:real_world_totals} gives us an insight into how each of the selected approaches perform on real-world binaries.
We see that Miasm has the highest \fscore score estimation, but as we have seen from Table~\ref{tab:real_world_static_dfg_times} it also has scalability issues with respect to memory consumption.
We see that angr misses a significant number of data flows and also reports a very large number of assumed false positives.
We use this as an opportunity to improve the state of the art in data-flow analysis.

\begin{table} 
\begin{center}
\scriptsize
\caption{The number of data flows discovered dynamically per alias class and the number of these discovered statically. Additionally, we show the number of data flows discovered statically only.}
\label{tab:real_world_performance}
\scalebox{0.85}{
\begin{tabular}{ l r r r r r r r}
 Alias Class & Dyn & angr & angr \% & Ghidra & Ghidra \% & Miasm & Miasm \%  \\
 \hline
 \multicolumn{8}{c}{chmod} \\\hline
(G, G) & $16$  & $0$  & $0.00\%$ & $0$ & $0.00\%$ & $0$ & $0.00\%$\\
(H, H)     & $2$  & $0$ & $0.00\%$  & $0$ & $0.00\%$ & $1$ & $50.00\%$\\
(S, S)   & $97$  & $24$ & $24.74\%$ & $2$ & $2.06\%$ & $4$ & $6.19\%$\\
unknown         & $3$ & $1$ & $33.33\%$ & $0$ & $0.00\%$ &  $0$& $0.00\%$ \\
Static-only & - & $1,331$ & - & $363$ & - & $22$ & - \\
\hline
 \multicolumn{8}{c}{cp} \\\hline
 (F, F) & $4$ & $0$ & $0.00\%$ & $0$ & $0.00\%$ & $0$ & $0.00\%$\\
(G, G)   & $3$  & $0$ & $0.00\%$ & $0$ & $0.00\%$ & $0$ & $0.00\%$\\
(H, H)       & $2$  & $0$ & $0.00\%$ & $0$ & $0.00\%$ & $0$ & $0.00\%$\\
(S, S)     & $511$  & $128$ & $25.05\%$ & $9$ & $1.76\%$ & $13$ & $2.54\%$\\
unknown           & $11$  & $2$ & $18.18\%$ & $2$ & $18.18\%$ & $0$ & $0.00\%$\\
Static-only & - & $1,555$ & - & $654$ & - & $56$ & - \\
\hline
 \multicolumn{8}{c}{ls} \\\hline
(F, F) & $1$ & $0$ & $0.00\%$ & $0$ & $0.00\%$ & $0$ & $0.00\%$\\
(G, G)   & $66$ & $7$ & $10.61\%$ & $4$ & $6.06\%$ & $0$ & $0.00\%$\\
(H, H)       & $1$ & $0$ & $0.00\%$ & $0$ & $0.00\%$ & $0$ & $0.00\%$\\
(S, S)     & $208$ & $96$ & $46.15\%$ & $10$ & $4.81\%$ & $37$ & $17.79\%$\\
unknown           & $10$ & $1$ & $10.00\%$ & $0$ & $0.00\%$ & $1$ & $10.00\%$\\
Static-only & - & $3,030$ & - & $572$ & - & $94$ & - \\
 \hline
 \multicolumn{8}{c}{Apache-Httpd} \\\hline
(F, F) & $10$ & $9$ & $90.00\%$ & $0$ & $0.00\%$ & $9$ & $90.00\%$\\
(G, G)   & $11$  & $8$ & $72.73\%$ & $0$ & $0.00\%$ & $0$ & $0.00\%$\\
(H, H)       & $1$ & $1$ & $100.00\%$ & $0$ & $0.00\%$ & $1$ & $100.00\%$\\
(S, S)     & $188$ & $133$ & $70.74\%$ & $19$ & $10.11\%$ & $188$ & $100.00\%$\\
unknown           & $2$ & $1$    & $50.00\%$ & $0$      & $0.00\%$ &  $0$ & $0.00\%$\\
Static-only & - & $125$ & - & $70$ & - & $36$ & - \\
\hline
 \multicolumn{8}{c}{Mujs} \\\hline
(F, F) & $8$ & $2$    & $25.00\%$  & $0$      & $0.00\%$ & $2$     & $25.00\%$\\
(S, S)     & $1,333$ & $519$  & $38.93\%$ & $25$ & $1.88\%$    & $217$ & $16.28\%$\\
unknown           & $6$ & $3$     & $50.00\%$ & $0$      & $0.00\%$ & $0$     & $0.00\%$\\
Static-only & - & $976$ & - & $363$ & - & $57$ & - \\
\hline
 \multicolumn{8}{c}{CJson} \\\hline
(S, S) & $90$  & $79$ & $87.78\%$ & $7$     & $7.78$ & $90$   & $100.00\%$\\
Static-only & - & $70$ & - & $31$ & - & $26$ & -
\end{tabular}
}
\end{center}

\end{table}

\begin{table} 
\begin{center}
\scriptsize
\caption{Performance of selected approaches over all target real-world binaries.}
\label{tab:real_world_totals}
\begin{tabular}{ l r r r}
 & angr & Ghidra & Miasm  \\ 
True positives (lower bound) & $1,014$ & $78$ & $563$  \\
False positives (upper bound) & $7,087$ & $2,053$ & $291$ \\
False negatives (lower bound) & $1,570$ & $2,506$ & $2,021$ \\
\hline
Precision (lower bound) & $0.1252$ & $0.0366$ & $0.6593$ \\
Recall (estimation) & $0.3924$ & $0.0302$ & $0.2179$ \\
\fscore score (estimation) & $0.1898$ & $0.0331$ & $0.3275$
\end{tabular}
\end{center}
\end{table} \mclearpage
\section{Improving the state of the art}
\label{sec:improving_the_state_of_the_art}
In this section, we leverage the results of our evaluation to improve the state of the art in static data-flow analysis.
For this purpose, we select angr as the data-flow analysis implementation to investigate and improve.
In Section~\ref{sec:room_for_improvement}, we highlight specific behavior of angr that reveals opportunities for improvement.
In Section~\ref{sec:data_flow_model_extensions} we introduce three novel data-flow model extensions that serve as alternatives to angr's behavior and show in Section~\ref{sec:evaluating_model_extensions} that these do indeed yield better results.
Finally, in Section~\ref{sec:security_application}, we show that leveraging these model extensions in a vulnerability-discovery context leads to an improved recovery of vulnerability-related instructions.

\subsection{Identifying improvement opportunities}
\label{sec:room_for_improvement}
In our evaluation of the selected approaches, we see in Table~\ref{tab:real_world_totals} that angr has a significant number of false negatives and assumed false positives.
This matches what we see in Table~\ref{tab:microbenchmark_evaluation}, where angr reports many impossible data flows (false positives).
These false positives and negatives are a clear indication that angr can be improved by fine-tuning its restrictions for when to report data flows.

In order to show the specific improvement opportunities, we present two additional categorizations of our microbenchmarks in Tables~\ref{tab:microbenchmark_callee_slice} and~\ref{tab:microbenchmark_offset_slice}.
Table~\ref{tab:microbenchmark_callee_slice} shows all unconditional data flows, paired with their underspecified counterparts, where the target data flow is interrupted by a function call.
We show an example of such a pair of data flows in Listings~\ref{list:unconditional_no_callee} and~\ref{list:underspecified_callee}.
From Table~\ref{tab:microbenchmark_callee_slice}, we clearly see that the presence of a callee function causes angr to not report the target data flow.
Since the callee function introduces out-of-scope modifications to the program state, the ground truth is underspecified.
Therefore, one could argue that disrupting all data flows that cross the callee function is an acceptable approach for an intra-procedural data-flow analysis to take.
However, in Section~\ref{sec:data_flow_model_extensions} we propose an alternative approach and in Section~\ref{sec:evaluating_model_extensions} we show that this reflects real-world behavior more accurately and therefore yields better results.

\begin{table} 
\begin{center}
\scriptsize
\caption{The change (in boldface) introduced by \modelextensioncallee and \modelextensionreturn in how angr reports data flows interrupted by a callee function.}
\label{tab:microbenchmark_callee_slice}
\begin{tabular}{ l l l | r r | r r }
& & & \multicolumn{2}{c}{angr} & \multicolumn{2}{|c}{\angrc}\\
 Alias Class & Ground Truth & Callee & Edge  & Edge \% & Edge  & Edge \%  \\ 
(F, F) & Unconditional  & No & 158 & $100.00\%$ & 158 & $100.00\%$ \\
(F, F) & Under-specified & Yes & $0$ & $0.00\%$ & $\mathbf{180}$  & $\mathbf{100.00\%}$ \\
(G, G) & Unconditional  & No & 170 & $100.00\%$ & 170 & $100.00\%$\\
(G, G) & Under-specified & Yes & $0$ & $0.00\%$ & $\mathbf{168}$ & $\mathbf{100.00\%}$\\
(H, H) & Unconditional & No & 376 & $100.00\%$ & 376 & $100.00\%$\\
(H, H) & Under-specified & Yes & $0$ & $0.00\%$ & $\mathbf{376}$  & $\mathbf{100.00\%}$\\
(S, S) & Unconditional & No & 115 & $100.00\%$ & 115 & $100.00\%$ \\
(S, S) & Under-specified & Yes & $0$ & $0.00\%$ & $\mathbf{135}$ & $\mathbf{100.00\%}$
\end{tabular}
\end{center}
\end{table}

\begin{minipage}[t]{0.90\columnwidth}
\begin{lstlisting}[language={[x64]Assembler}, caption={A fully-specified unconditional data flow between exists between lines~\ref{list:unconditional_no_callee:write} and~\ref{list:unconditional_no_callee:read}.}, label={list:unconditional_no_callee}, escapechar=|, boxpos=t]
mov BYTE PTR [rsp-0x1],dil ; Write|\label{list:unconditional_no_callee:write}|
mov al,BYTE PTR [rsp-0x1]  ; Read|\label{list:unconditional_no_callee:read}|
\end{lstlisting}
\end{minipage}

\begin{minipage}[t]{0.90\columnwidth}
\begin{lstlisting}[language={[x64]Assembler}, caption={The counterpart of Listing~\ref{list:unconditional_no_callee}. The data flow is interrupted by a function call (line~\ref{list:underspecified_callee:call}), resulting in a underspecified data flow between lines~\ref{list:underspecified_callee:write} and~\ref{list:underspecified_callee:read}.}, label={list:underspecified_callee}, escapechar=|, boxpos=t]
sub  rsp,0x10
mov  BYTE PTR [rsp+0xf],dil ; Write|\label{list:underspecified_callee:write}|
call 11e9|\label{list:underspecified_callee:call}|
mov  al,BYTE PTR [rsp+0xf]  ; Read|\label{list:underspecified_callee:read}|
add  rsp,0x10
\end{lstlisting}
\end{minipage}

We highlight our second opportunity for improvement with Table~\ref{tab:microbenchmark_offset_slice}.
In this table, we select all fully-specified data flows and categorize these with respect to ground truth, and whether or not an offset transformation was applied to both the write and read pointer with equal offsets (discussed in Section~\ref{sec:microbenchmark_test_cases}).
We show an example of such a test case in Listing~\ref{list:impossible_distinct_offsets}.
In Table~\ref{tab:microbenchmark_offset_slice}, we see that angr reports fully-specified, impossible data flows with distinct offsets.
Since these are fully-specified test cases, we can confirm angr's behavior as erroneous.
In order to understand how to correct this behavior, we investigate angr's source code.
Shortly, angr exhibits this behavior due to how it treats undefined memory addresses.
Such an address is artificially concretized to a constant specified in angr's source code.
This effectively assumes all undefined memory addresses alias.
We propose an alternative approach in Section~\ref{sec:data_flow_model_extensions} and in Section~\ref{sec:evaluating_model_extensions} we show that this reflects real-world behavior more accurately and therefore yields better results.

\begin{minipage}[t]{0.95\columnwidth}
\begin{lstlisting}[language={[x64]Assembler}, caption={A fully-specified impossible data flow exists between instructions~\ref{list:impossible_distinct_offsets:write} and~\ref{list:impossible_distinct_offsets:read} due to the distinct offsets of register \texttt{rsi} that are accessed: \texttt{[rsi+0x0]} vs. \texttt{[rsi+0x1]}. }, label={list:impossible_distinct_offsets}, escapechar=|, boxpos=t]
lea    rax,[rsi+0x1]
mov    QWORD PTR [rsp-0x8],rax
mov    BYTE PTR [rsi],dil      ; Write|\label{list:impossible_distinct_offsets:write}|
mov    rax,QWORD PTR [rsp-0x8]
mov    al,BYTE PTR [rax]       ; Read|\label{list:impossible_distinct_offsets:read}|
\end{lstlisting}
\end{minipage}

\begin{table} 
\begin{center}
\scriptsize
\caption{The change (in boldface) introduced by \modelextensionf with respect to how angr reports data flows between pointers transformed by equal or distinct offsets.}
\label{tab:microbenchmark_offset_slice}
\scalebox{0.85}{
\begin{tabular}{ l l l | r r | r r }
& & & \multicolumn{2}{c}{angr} & \multicolumn{2}{|c}{\angrf}\\
 Alias Class & Ground Truth & Equal Offset & Edge  & Edge \% & Edge  & Edge \%  \\ 
(F, F) & Unconditional  & Yes & $158$ & $100.00\%$ & $158$ & $100.00\%$  \\
(F, F) & Impossible & No & $72$ & $100.00\%$ & $\mathbf{0}$ & $\mathbf{0.00\%}$ \\
(G, G) & Unconditional  & Yes & 170 & $100.00\%$ & 170 & $100.00\%$ \\
(G, G) & Impossible & Yes &$0$ & $0.00\%$ & $0$ & $0.00\%$ \\
(G, G) & Impossible & No & $0$ & $0.00\%$ & $0$ & $0.00\%$ \\
(H, H) & Unconditional & Yes & $376$ & $100.00\%$ & $376$ & $100.00\%$ \\
(H, H) & Impossible & Yes & $8,480$ & $100.00\%$ & $\mathbf{3,402}$ & $\mathbf{40.12\%}$\\
(H, H) & Impossible & No & $4,508$ & $100.00\%$ & $\mathbf{475}$ & $\mathbf{10.54\%}$\\
(S, S) & Unconditional & Yes &  $115$ & $100.00\%$  & $115$ & $100.00\%$ \\
(S, S) & Impossible & Yes & $717$ & $38.88\%$ & $\mathbf{0}$ & $\mathbf{0.00\%}$ \\
(S, S) & Impossible & No & $0$ & $0.00\%$ & $0$ & $0.00\%$
\end{tabular}
}
\end{center}
\end{table}

\subsection{Data-flow model extensions}
\label{sec:data_flow_model_extensions}
Our novel model extensions focus on addressing the limitations discussed in Section~\ref{sec:room_for_improvement}.
Two extensions introduce more precise handling of the state of a caller function upon return from a callee function, discussed in Section~\ref{sec:callee_function_summarization}.
The third model extension introduces a simple, but effective, way to improve field sensitivity of static data-flow analysis, as described in Section~\ref{sec:field_sensitivity}.
We show how these three model extensions vastly improve the accuracy of static data-flow analysis in Section~\ref{sec:improving_the_state_of_the_art}.

\subsubsection{Handling function calls}
\label{sec:callee_function_summarization} 
\paragraph{\modelextensioncallee: Leveraging calling convention}
A calling convention defines how function parameters and return values are passed between a caller function and its callee functions.
We propose a model extension that improves intra-procedural data-flow analysis by incorporating information about the calling convention implemented by the target function. 
Consider Listing~\ref{list:callee_heuristic_1}, in which register \texttt{rdi} is used to pass a memory address from the caller function \texttt{f\_target} (line~\ref{list:callee_heuristic_1:arg}) to callee function \texttt{f\_callee} (line~\ref{list:callee_heuristic_1:callee}) as a function argument.
By identifying such function arguments, a policy can be used to determine whether to preserve or kill definitions at these addresses.
We implement our model extensions \modelextensioncallee with a policy that naively assumes callee functions have no impact on intra-procedural data flows and therefore data flows should be preserved.

\paragraph{\modelextensionreturn:  Stack frame preservation}
Conventionally, callee functions are implemented to preserve and restore the stack frame of their caller function.
As the callee function is out of scope for an intra-procedural analysis, our model extension preserves the stack frame artificially for all callee functions.
This is challenging as it requires suppressing the effect that the \texttt{call} instruction itself has on the stack frame.
In architectures such as \texttt{x86\_64}, the \texttt{call} instruction pushes the subsequent instruction address to the stack automatically, modifying the stack pointer.
The matching return instruction, popping this instruction address and restoring the stack pointer, is out of scope of analysis.
Therefore, artificial restoration is necessary.


\subsubsection{Field sensitivity}
\label{sec:field_sensitivity}
\paragraph{\modelextensionf: Constant-based Field Disunion}
Programming languages often allow the programmer to group related data together into a structure (aka \emph{structs}). Separate values in such a struct are referred to as \emph{fields}.
In machine code, this is usually implemented by storing a \textit{base address} of the struct, and accessing the fields by computing an offset from the base address.
The sizes of the different fields must be defined at compile time, meaning the offsets to the different fields are constant and can be observed in the assembly code.
Our extension assumes there is no data flow between two memory access instructions using distinct constant offsets, as these represent different fields.

In Section~\ref{sec:room_for_improvement}, we mentioned that angr handles undefined memory addresses by concretizing them to a single arbitrary address.
We implement \modelextensionf such that instead of concretizing the entire undefined address, we only concretize the undefined registers used in the address expression.
This keeps address concretization sensitive to offsets, as required by \modelextensionf.

\subsection{Evaluating our model extensions}
\label{sec:evaluating_model_extensions}
We refer to angr extended with \modelextensioncallee and \modelextensionreturn as \angrc, with \modelextensionf as \angrf and with all three extensions as \angrcf.

We show the difference between angr and \angrc using our microbenchmarks in Table~\ref{tab:microbenchmark_callee_slice} and similarly for \angrf in Table~\ref{tab:microbenchmark_offset_slice}.
Table~\ref{tab:microbenchmark_callee_slice} shows that we have successfully extended angr to preserve data flows that cross a callee function.
Table~\ref{tab:microbenchmark_offset_slice} shows that we have reduced the cases where angr reports impossible data flows when two distinct offsets are employed.
An exception here is with the \texttt{(Heap, Heap)} alias class in which $475$ ($10.54\%$) impossible data flows with distinct offsets are still reported.
After manually investigating a number of the remaining cases, we concluded that the reason for this is because of multi-byte memory accesses.
A multi-byte memory access instruction writes or reads to memory at a small range of addresses.
Even though two distinct offsets are used for the memory write and read instruction, the ranges overlap.
We also observe an unexpected improvement gained by \angrf reducing the reported impossible data flows with \emph{equal} offset transformations in the \texttt{(Heap, Heap)} and \texttt{(Stack, Stack)} alias classes.
We established that the reason for this is a secondary offset introduced by the compiler due to different data types accessed by the write pointer and read pointer.
Since \angrf does not distinguish between offsets added explicitly in the source code or by the compiler, it correctly does not report the data flow.

We prove the real-world improvement of \angrcf by re-performing our evaluation on real-world test cases and show the results in Table~\ref{tab:real_world_performance_angrc1c2f1}.
We see that \angrcf has a higher \fscore score estimation than angr, Ghidra and Miasm.
We gain a significant increase in true positives and reduction in assumed false positives.
Indeed, \angrcf achieves nearly perfect recall, meaning any real data flow is likely to be reported by \angrcf with near guaranteed certainty.
This is achieved, while simultaneously improving precision  from 13\% to 32\%, i.e., while reducing false positives.

\begin{table} %

\begin{center}
\scriptsize
\caption{The concrete improvement gained by extending angr with our model extensions \modelextensioncallee, \modelextensionreturn and \modelextensionf.}
\label{tab:real_world_performance_angrc1c2f1}
\begin{tabular}{ l r r}
 & angr & \angrcf  \\ 
True positives (lower bound) & $1,014$ & $2,569$  \\
False positives (upper bound) & $7,087$ & $5,351$ \\
False negatives (lower bound) & $1,570$ & $15$  \\
\hline
Precision (lower bound) & $0.1252$ & $0.3244$  \\
Recall (estimation) & $0.3924$ & $0.9942$  \\
\fscore score (estimation) & $0.1898$ & $0.4891$
\end{tabular}
\end{center}
\end{table}

\subsection{Security impact of our model extensions}
\label{sec:security_application}
Data flow is a critical component of vulnerability discovery. Many vulnerability discovery tools, such as BVdetector~\cite{DBLP:journals/infsof/TianXL20} and BinHunter~\cite{DBLP:conf/acsac/ArastehMRH24}, rely on data flow to capture binary instructions relevant to the vulnerability.
Improving the accuracy of flow analysis can help identify the relevant instructions more accurately and, in turn, help improve vulnerability discovery.
In this section, we present three case studies.
In each case study, we perform manual analysis on the source and assembly code of a vulnerable program in order to identify which instructions are relevant to the vulnerability in question, and to identify the (ground truth) data flows between these instructions.
We select these examples to show that \angrcf is capable of identifying the data flows between vulnerability-relevant instructions more accurately than the original angr.

\subsubsection{CVE-2018-5785}
This is an integer-overflow vulnerability that occurred in \texttt{openjpeg2} in the module responsible for processing bitmap image files.
The bitmap image file format includes a device-independent bitmap (DIB) header, which contains metadata about the image, for example its width, height and color format.
The first 4 bytes of this header specify the size of the header.
When this size is larger than $56$ bytes, the header includes a section for bitmasks, which are used to define which bits in a 32-bit pixel value correspond to the red, green, blue and alpha (transparency) channels. 
Listing~\ref{list:bitmast_source_code} shows the part of the source code of function \texttt{bmp\_read\_info\_header} in the vulnerable module, responsible for parsing the bitmasks.
In this code snippet, if the header size is at least 56 bytes, the program reads four bytes from the input file and combines them to reconstruct the red-color bitmask value. 
This is done by shifting and merging each byte into a 32-bit integer.
This code is the root cause of an integer overflow, because it allows the mask values to be zero.
This zero value is later used to calculate a negative shift value of a left-shift operator (\lstinline[language=C]{1 << -1}).
This is undefined behavior and may cause crashes or security issues.
As shown in Listing~\ref{list:bitmast_source_code}, the value of \texttt{biRedMask} is used as the left-hand operand in a subsequent operation, creating a data flow across lines 3, 4, 5, and 6. 
Listing~\ref{list:data_flow_2018} presents the corresponding assembly instructions for this code snippet. 
The value of \texttt{biRedMask}, stored in memory at \texttt{[RAX+0x28]}, is used in instruction \texttt{fb0d} (line~\ref{list:data_flow_2018:read-1}), creating a data flow between instructions \texttt{faf5} (line~\ref{list:data_flow_2018:write-1}) and \texttt{fb0d} (line~\ref{list:data_flow_2018:read-1}).
Similarly, data flows exist between instruction pairs \texttt{(fb16, fb2e)} and \texttt{(fb37, fb4f)}.
While the original angr fails to detect any of these data flows, \angrcf successfully captures all of them.
This behavior is consistent across other mask values as well, including \texttt{biGreenMask}, \texttt{biBlueMask}, and \texttt{biAlphaMask}.
In total \angrcf was able to recover 12 additional true positive data flows between vulnerability-related instructions.

\begin{center}
\noindent\begin{minipage}{0.90\columnwidth}
\begin{lstlisting}[
  language=C,
  caption={Part of source code function bmp\_read\_info\_header with integer overflow vulnerability},
  label={list:bitmast_source_code},
  breaklines=false,
  basicstyle=\ttfamily\scriptsize,
  columns=fullflexible,
  frame=single,
  numbers=left,
  numberstyle=\tiny,
  showstringspaces=false
]
if (header->biSize >= 56U) {

    header->biRedMask  = (OPJ_UINT32)getc(IN);
    header->biRedMask |= (OPJ_UINT32)getc(IN) << 8;
    header->biRedMask |= (OPJ_UINT32)getc(IN) << 16;
    header->biRedMask |= (OPJ_UINT32)getc(IN) << 24;
}
\end{lstlisting}
\end{minipage}
\end{center}

\begin{center}
\noindent\begin{minipage}{0.90\columnwidth}
\begin{lstlisting}[language={[x64]Assembler}, caption={Corresponding assembly instructions for Listing~\ref{list:bitmast_source_code}, which contains an integer overflow}, label={list:data_flow_2018}, escapechar=|]
fae3: MOV  RAX,qword ptr [RBP + local_10]
fae7: MOV  RDI,RAX
faea: CALL <EXTERNAL>::_IO_getc
faef: MOV  EDX,EAX
faf1: MOV  RAX,qword ptr [RBP + local_18]
faf5: MOV  dword ptr [RAX + 0x28],EDX     ; Write-1|\label{list:data_flow_2018:write-1}|   
faf8: MOV  RAX,qword ptr [RBP + local_10]
fafc: MOV  RDI,RAX
faff: CALL <EXTERNAL>::_IO_getc            
fb04: SHL  EAX,0x8
fb07: MOV  EDX,EAX
fb09: MOV  RAX,qword ptr [RBP + local_18]
fb0d: MOV  EAX,dword ptr [RAX + 0x28]     ; read-1|\label{list:data_flow_2018:read-1}|
fb10: OR   EDX,EAX
fb12: MOV  RAX,qword ptr [RBP + local_18]
fb16: MOV  dword ptr [RAX + 0x28],EDX	   ; Write-2|\label{list:data_flow_2018:write-2}|
fb19: MOV  RAX,qword ptr [RBP + local_10]
fb1d: MOV  RDI,RAX
fb20: CALL <EXTERNAL>::_IO_getc              
fb25: SHL  EAX,0x10
fb28: MOV  EDX,EAX
fb2a: MOV  RAX,qword ptr [RBP + local_18]
fb2e: MOV  EAX,dword ptr [RAX + 0x28]     ; read-2|\label{list:data_flow_2018:read-2}|
fb31: OR   EDX,EAX
fb33: MOV  RAX,qword ptr [RBP + local_18]
fb37: MOV  dword ptr [RAX + 0x28],EDX     ; Write-3|\label{list:data_flow_2018:write-3}|
fb3a: MOV  RAX,qword ptr [RBP + local_10]
fb3e: MOV  RDI,RAX
fb41: CALL <EXTERNAL>::_IO_getc             
fb46: SHL  EAX,0x18
fb49: MOV  EDX,EAX
fb4b: MOV  RAX,qword ptr [RBP + local_18]
fb4f: MOV  EAX,dword ptr [RAX + 0x28]     ; read-3|\label{list:data_flow_2018:read-3}|
fb52: OR   EDX,EAX
\end{lstlisting}
\end{minipage}
\end{center}

\subsubsection{CVE-2022-4904}
This is a buffer overflow vulnerability that occurred in the \texttt{c-ares} package, a C library designed for asynchronous DNS (Domain Name System) resolution.
This vulnerability exists in the function \texttt{config\_sortlist}, responsible for parsing and storing the sortlist configuration used in DNS resolution.
We show a snippet of this function to illustrate the vulnerability in Listing~\ref{list:ares_source_code}.
In this listing, the while loops on lines~\ref{list:ares_source_code:while1:start}-\ref{list:ares_source_code:while1:end} and~\ref{list:ares_source_code:while2:start}-\ref{list:ares_source_code:while2:end} are responsible for extracting an IP address or range from the configuration file.
In line 5, the \texttt{memcpy} function copies \lstinline[language=C]{q - s} bytes from this IP address in the \texttt{str} buffer to the \texttt{ipbuf} buffer.
Note that no check is placed on the number of bytes copied into the \texttt{ipbuf} buffer.
Therefore, if the number of bytes copied exceeds the size of \texttt{ipbuf} (16 bytes), a buffer overflow occurs.
A similar issue arises on line 14 with the \texttt{ipbufpfx} buffer. 
Tracking the values of \texttt{q} and \texttt{str} is crucial for detecting these overflows, as they determine whether the buffer boundaries are respected in lines 5 and 14.

Listing~\ref{list:data_flow_2022} shows part of the corresponding assembly instructions of the function \texttt{config\_sortlist}. The values of \texttt{str} and \texttt{q} used in the \texttt{size} function argument of \texttt{memcpy} are first defined at addresses \texttt{bb96} (line~\ref{list:data_flow_2022:write-1}) and \texttt{bba9} (line~\ref{list:data_flow_2022:write_local_10}). The value of \texttt{str} has been read in \texttt{bc00} (line~\ref{list:data_flow_2022:add_local_90}) and \texttt{bc0a} (line~\ref{list:data_flow_2022:add_local_90_second}) as arguments for the first \texttt{memcpy} function and then again read at addresses \texttt{bc93} (line~\ref{list:data_flow_2022:read-1}) and \texttt{bc9d} (line~\ref{list:data_flow_2022:add_local_90_third}) for the second \texttt{memcpy} function, and remains constant in between. This creates a data flow between (\texttt{bb96}, \texttt{bc00}), (\texttt{bb96}, \texttt{bc0a}), (\texttt{bb96}, \texttt{bc93}), and (\texttt{bb96}, \texttt{bc9d}). The original angr fails to identify the last two data flows blocked by the \texttt{memcpy} function, while \angrcf identifies all of them correctly. Therefore, relying solely on angr’s data-flow analysis may lead to missing critical instructions related to the vulnerability. Our evaluation demonstrated that \angrcf was able to recover 36 true positive data flows involving vulnerable instructions that were missed by the original angr.

\begin{center}
\noindent\begin{minipage}{0.90\columnwidth}
\begin{lstlisting}[
  language=C,
  caption={A small part of the source code of the function \texttt{config\_sortlist}},
  label={list:ares_source_code},
  breaklines=false,
  basicstyle=\ttfamily\scriptsize,
  columns=fullflexible,
  frame=single,
  numbers=left,
  numberstyle=\tiny,
  showstringspaces=false,
  escapechar=@
]

q = str;
while (*q && *q != '/' && *q != ';' && !ISSPACE(*q))@\label{list:ares_source_code:while1:start}@
 q++;@\label{list:ares_source_code:while1:end}@
memcpy(ipbuf, str, q-str);
ipbuf[q-str] = '\0';
/* Find the prefix */

if (*q == '/')
{
const char *str2 = q+1;
while (*q && *q != ';' && !ISSPACE(*q))@\label{list:ares_source_code:while2:start}@
 q++;@\label{list:ares_source_code:while2:end}@
memcpy(ipbufpfx, str, q-str);
ipbufpfx[q-str] = '\0';
str = str2;
}
\end{lstlisting}
\end{minipage}
\end{center}

\begin{center}
\noindent\begin{minipage}{0.90\columnwidth}
\begin{lstlisting}[language={[x64]Assembler}, caption={Corresponding assembly instructions for function \texttt{config\_sortlist}}, label={list:data_flow_2022}, escapechar=|]
bb8e: MOV  qword ptr [RBP + local_80],RDI
bb92: MOV  qword ptr [RBP + local_88],RSI
bb96: MOV  qword ptr [RBP + local_90],RDX  ; Write-1|\label{list:data_flow_2022:write-1}|
bb9d: JMP  LAB_0010bf6c 
bba2: MOV  RAX,qword ptr [RBP + local_90]
bba9: MOV  qword ptr [RBP + local_10],RAX  |\label{list:data_flow_2022:write_local_10}|
bbad: JMP  LAB_0010bbb4                 
bbaf: ADD  qword ptr [RBP + local_10],0x1  |\label{list:data_flow_2022:add_local_10}|
...
bbf8: TEST  EAX,EAX
bbfa: JZ    LAB_0010bbaf         
bbfc: MOV   RAX,qword ptr [RBP + local_10]
bc00: SUB   RAX,qword ptr [RBP + local_90] |\label{list:data_flow_2022:add_local_90}|
bc07: MOV   RDX,RAX
bc0a: MOV   RCX,qword ptr [RBP + local_90] |\label{list:data_flow_2022:add_local_90_second}|
bc11: LEA   RAX=>local_58,[RBP + -0x50]
bc15: MOV   RSI,RCX
bc18: MOV   RDI,RAX
bc1b: CALL  <EXTERNAL>::memcpy|\label{list:data_flow_2022:memcpy1}|
...
bc4d: ADD   qword ptr [RBP + local_10],0x1 |\label{list:data_flow_2022:second_write_local_10}|
bc52: RAX,  qword ptr [RBP + local_10]
...
bc8f: MOV   RAX,qword ptr [RBP + local_10]|\label{list:data_flow_2022:read_local_10}| 
bc93: SUB   RAX,qword ptr [RBP + local_90] ; read-1|\label{list:data_flow_2022:read-1}|
bc9a: MOV   RDX,RAX
bc9d: MOV   RCX,qword ptr [RBP + local_90] |\label{list:data_flow_2022:add_local_90_third}|
bca4: LEA   RAX=>local_78,[RBP + -0x70]
bca8: MOV   RSI,RCX
bcab: MOV   RDI,RAX
bcae: CALL  <EXTERNAL>::memcpy|\label{list:data_flow_2022:memcpy2}|

\end{lstlisting}
\end{minipage}
\end{center}

\subsubsection{CVE-2023-31130}

This is a buffer underflow vulnerability in the \texttt{c-ares} library's \texttt{ares\_inet\_net\_pton()} function, which can be triggered by certain malformed IPv6 addresses. Due to the scope of the vulnerability, we only focus on a small segment, shown in Listing~\ref{list:ares_source_code_two}.
In this listing, the numerical value of an IPv6 hexadecimal segment should be stored in \texttt{val}.
However, the function fails to constrain \texttt{val} within the correct numerical bounds ($0 \leq \texttt{val} < 2^{16}$).
In order to discover the possible inappropriate value for \texttt{val}, it is essential to recover all instructions that contribute to computing this value.
Therefore, it is necessary to identify the data flow in Listing~\ref{list:ares_source_code_two} from line~\ref{list:ares_source_code_two:shift_val} to~\ref{list:ares_source_code_two:assign}.
Listing~\ref{list:data_flow_2023} shows the corresponding assembly instructions for the source code.
Observe that there should be a data flow between \texttt{(16608, 1661c)}. While the original angr does not report this data flow, \angrcf does.
In total, \angrcf discovers 27 more correct data flows related to this vulnerability than the original angr. 

\begin{center}
\noindent\begin{minipage}{0.90\columnwidth}
\begin{lstlisting}[
  language=C,
  caption={Part of source code function inet\_net\_pton\_ipv6 that contains buffer underflow vulnerability},
  label={list:ares_source_code_two},
  breaklines=false,
  basicstyle=\ttfamily\scriptsize,
  columns=fullflexible,
  frame=single,
  numbers=left,
  numberstyle=\tiny,
  showstringspaces=false,
  escapechar=@
]
  val <<= 4; @\label{list:ares_source_code_two:shift_val}@
  val |= aresx_sztoui(pch - xdigits); @\label{list:ares_source_code_two:assign}@
  if (++digits > 4)
    goto enoent;
\end{lstlisting}
\end{minipage}
\end{center}

\begin{center}
\noindent\begin{minipage}{0.90\columnwidth}
\begin{lstlisting}[language={[x64]Assembler}, caption={Corresponding assmbly instructions for function inet\_net\_pton\_ipv6 that contains buffer underflow vulnerability}, label={list:data_flow_2023}, escapechar=|]
16608: SHL  dword ptr [RBP + local_30],0x4    ; Write|\label{list:data_flow_2023:write}|
1660c: MOV  RAX,qword ptr [RBP + local_48]
16610: SUB  RAX,qword ptr [RBP + local_20]
16614: MOV  RDI,RAX
16617: CALL aresx_sztoui                         
1661c: OR   dword ptr [RBP + local_30],EAX   ; read|\label{list:data_flow_2023:read}|
\end{lstlisting}
\end{minipage}
\end{center} \mclearpage
\section{Future Work}
In Section~\ref{sec:pointer_transformation} we introduce the offset pointer transformation that operates by adding a value to the write or read pointer of a target data flow.
Additional pointer transformations can be implemented, such as killing (redefining) the definition of the write instruction.
Any additional transformation will yield more insight into how effectively a static data-flow analysis approach can identify data flows in case of such a transformation.

In Section~\ref{sec:alias_classes} we define a pointer origin for pointers passed as function arguments, the \emph{foreign pointers} and for pointers returned from a memory allocation function, the \emph{heap pointers}.
It is possible to bridge these two pointer origins with pointers returned from functions other than memory allocation functions.
Such a pointer is essentially also a type of foreign pointer, as no information is available regarding its definition site.

\jelena{Nicolaas, I would add Limitations to this section so it becomes "Limitations and Future Work". Then I would say that one limitation is that all approximations can lead to errors so the real challenge is identifying those that help more than hurt. Add any other limitations like we don't do inter-procedural. We also know that some inaccuracies come not from approximations per se but from internal models of code that are imprecise.}



 \mclearpage
\section{Related Work}
To the best of our knowledge, we are the first to evaluate static data-flow analysis approaches on an extensive data set of binary executables.
There have been a number of other benchmarks with related, but orthogonal goals.
Andriesse et al.~\cite{DBLP:conf/uss/AndriesseCVSB16} and similarly Pang et al.~\cite{DBLP:conf/uss/PangZYM022, DBLP:conf/sp/PangYCKPMX21} evaluate disassembler implementations on a data set consisting of binaries extracted from the SPEC CPU 2006 benchmark, as well as real-world binaries.
Such an evaluation has a number of overlapping goals with ours, such as establishing ground truth information for real-world binaries, but disassembly is a problem orthogonal to data-flow analysis.
Di Federico et al. evaluate CFG recovery by creating a data set of binaries with ground truth function boundaries~\cite{DBLP:conf/cc/FedericoPA17}.
They compare their novel approach REV.NG with other approaches toward function boundary detection.
Data-flow analysis involves a number of challenges independent of control-flow analysis, as discussed in Section~\ref{sec:scope}.\jelena{I think what you want to say here is that while we could have used these datasets they may not properly challenge data flow analysis, which is what our benchmarks do. }

Hind~\cite{DBLP:conf/paste/Hind01} has surveyed a number of approximating alias analysis solutions on source code.
This work is complementary to ours. It approaches the challenge from a theoretical perspective, while we focus on measuring the concrete strengths and weaknesses of implementations of binary data-flow analysis.

Machiry et al. introduced AutoFacts, an approach to inject synthetic facts into real-world programs~\cite{autofacts}.
These facts allow for ground truth knowledge, that is both sound and complete, regarding aliasing pointers.
The injected facts, however, are entirely separate from the logic of the program into which they are injected.
Our approach focuses on the other two ends of this spectrum: testing microbenchmarks that are disjoint from real-world program logic and testing data flows fully intertwined in real-world program logic.
Additionally, while the AutoFacts data set is introduced, authors did not use it to analyze the implementations of static program analysis.
In both cases~\cite{DBLP:conf/paste/Hind01, autofacts} the alias approximations are divided into a number of dimensions, called sensitivities.
Since our selected approaches (angr, Miasm, Ghidra) do not allow for enabling or disabling these sensitivities,  we do not use AutoFacts in our evaluation.\jelena{but it sounds like this is most closely related work, so is there something else we can say about it other than "it tests other things than what we care about"} \mclearpage
\section{Conclusion}
In this paper, we introduced a novel approach to classify data flows, namely alias classes.
Using these alias classes as a guide, we implemented an open source framework to create a data set of both microbenchmarks and real-world binaries to evaluate data-flow analysis implementations.
We also implement an open source framework to perform this evaluation.
Finally, we evaluate angr, Ghidra and Miasm using our data set and framework and provide insights into limitations that each engine has with regards to data flow analysis.
By leveraging this evaluation, we propose three model extensions to angr that greatly improve accuracy of its  data-flow analysis and can be used to improve vulnerability discovery. \mclearpage

\bibliographystyle{IEEEtran}
\bibliography{main}

\begin{thebibliography}{10}
\providecommand{\url}[1]{#1}
\csname url@samestyle\endcsname
\providecommand{\newblock}{\relax}
\providecommand{\bibinfo}[2]{#2}
\providecommand{\BIBentrySTDinterwordspacing}{\spaceskip=0pt\relax}
\providecommand{\BIBentryALTinterwordstretchfactor}{4}
\providecommand{\BIBentryALTinterwordspacing}{\spaceskip=\fontdimen2\font plus
\BIBentryALTinterwordstretchfactor\fontdimen3\font minus \fontdimen4\font\relax}
\providecommand{\BIBforeignlanguage}[2]{{%
\expandafter\ifx\csname l@#1\endcsname\relax
\typeout{** WARNING: IEEEtran.bst: No hyphenation pattern has been}%
\typeout{** loaded for the language `#1'. Using the pattern for}%
\typeout{** the default language instead.}%
\else
\language=\csname l@#1\endcsname
\fi
#2}}
\providecommand{\BIBdecl}{\relax}
\BIBdecl

\bibitem{DBLP:journals/toplas/BalakrishnanR10}
\BIBentryALTinterwordspacing
G.~Balakrishnan and T.~W. Reps, ``{WYSINWYX:} what you see is not what you execute,'' \emph{{ACM} Trans. Program. Lang. Syst.}, vol.~32, no.~6, pp. 23:1--23:84, 2010. [Online]. Available: \url{https://doi.org/10.1145/1749608.1749612}
\BIBentrySTDinterwordspacing

\bibitem{angr}
{angr}, ``{The Angr binary analysis platform},'' \url{http://angr.io}, 2016.

\bibitem{ghidra}
{Ghidra}, ``{Ghidra},'' \url{https://ghidra-sre.org/}, 2022.

\bibitem{miasm}
{Miasm}, ``{Miasm},'' \url{https://miasm.re}, 2019.

\bibitem{DBLP:journals/toplas/Reps00}
\BIBentryALTinterwordspacing
T.~W. Reps, ``Undecidability of context-sensitive data-independence analysis,'' \emph{{ACM} Trans. Program. Lang. Syst.}, vol.~22, no.~1, pp. 162--186, 2000. [Online]. Available: \url{https://doi.org/10.1145/345099.345137}
\BIBentrySTDinterwordspacing

\bibitem{DBLP:books/aw/AhoSU86}
\BIBentryALTinterwordspacing
A.~V. Aho, R.~Sethi, and J.~D. Ullman, \emph{Compilers: Principles, Techniques, and Tools}, ser. Addison-Wesley series in computer science / World student series edition.\hskip 1em plus 0.5em minus 0.4em\relax Addison-Wesley, 1986. [Online]. Available: \url{https://www.worldcat.org/oclc/12285707}
\BIBentrySTDinterwordspacing

\bibitem{DBLP:conf/scam/KissJLG03}
\BIBentryALTinterwordspacing
{\'{A}}.~Kiss, J.~J{\'{a}}sz, G.~Lehotai, and T.~Gyim{\'{o}}thy, ``Interprocedural static slicing of binary executables,'' in \emph{3rd {IEEE} International Workshop on Source Code Analysis and Manipulation {(SCAM} 2003), 26-27 September 2003, Amsterdam, The Netherlands}.\hskip 1em plus 0.5em minus 0.4em\relax {IEEE} Computer Society, 2003, p. 118. [Online]. Available: \url{https://doi.org/10.1109/SCAM.2003.1238038}
\BIBentrySTDinterwordspacing

\bibitem{DBLP:conf/vstte/BalakrishnanRMT05}
\BIBentryALTinterwordspacing
G.~Balakrishnan, T.~W. Reps, D.~Melski, and T.~Teitelbaum, ``{WYSINWYX:} what you see is not what you execute,'' in \emph{Verified Software: Theories, Tools, Experiments, First {IFIP} {TC} 2/WG 2.3 Conference, {VSTTE} 2005, Zurich, Switzerland, October 10-13, 2005, Revised Selected Papers and Discussions}, ser. Lecture Notes in Computer Science, B.~Meyer and J.~Woodcock, Eds., vol. 4171.\hskip 1em plus 0.5em minus 0.4em\relax Springer, 2005, pp. 202--213. [Online]. Available: \url{https://doi.org/10.1007/978-3-540-69149-5\_22}
\BIBentrySTDinterwordspacing

\bibitem{rice1953classes}
H.~G. Rice, ``Classes of recursively enumerable sets and their decision problems,'' \emph{Transactions of the American Mathematical society}, vol.~74, no.~2, pp. 358--366, 1953.

\bibitem{DBLP:journals/toplas/Ramalingam94}
\BIBentryALTinterwordspacing
G.~Ramalingam, ``The undecidability of aliasing,'' \emph{{ACM} Trans. Program. Lang. Syst.}, vol.~16, no.~5, pp. 1467--1471, 1994. [Online]. Available: \url{https://doi.org/10.1145/186025.186041}
\BIBentrySTDinterwordspacing

\bibitem{pyelftools}
{pyelftools}, ``{Pyelftools},'' \url{https://github.com/eliben/pyelftools}, 2023.

\bibitem{coreutils}
{coreutils}, ``{Coreutils - GNU core utilities},'' \url{https://www.gnu.org/software/coreutils/}, 2023.

\bibitem{httpd}
{apache}, ``{Apache - HTTP Server Project},'' \url{https://httpd.apache.org/}, 2023.

\bibitem{mujs}
{Artifex}, ``{MuJS},'' \url{https://mujs.com/}, 2023.

\bibitem{cjson}
{cjson}, ``{cJSON - Ultralightweight JSON parser in ANSI C },'' \url{https://github.com/DaveGamble/cJSON}, 2023.

\bibitem{DBLP:conf/pldi/LukCMPKLWRH05}
\BIBentryALTinterwordspacing
C.~Luk, R.~S. Cohn, R.~Muth, H.~Patil, A.~Klauser, P.~G. Lowney, S.~Wallace, V.~J. Reddi, and K.~M. Hazelwood, ``Pin: building customized program analysis tools with dynamic instrumentation,'' in \emph{Proceedings of the {ACM} {SIGPLAN} 2005 Conference on Programming Language Design and Implementation, Chicago, IL, USA, June 12-15, 2005}, V.~Sarkar and M.~W. Hall, Eds.\hskip 1em plus 0.5em minus 0.4em\relax {ACM}, 2005, pp. 190--200. [Online]. Available: \url{https://doi.org/10.1145/1065010.1065034}
\BIBentrySTDinterwordspacing

\bibitem{DBLP:journals/infsof/TianXL20}
\BIBentryALTinterwordspacing
J.~Tian, W.~Xing, and Z.~Li, ``Bvdetector: {A} program slice-based binary code vulnerability intelligent detection system,'' \emph{Inf. Softw. Technol.}, vol. 123, p. 106289, 2020. [Online]. Available: \url{https://doi.org/10.1016/j.infsof.2020.106289}
\BIBentrySTDinterwordspacing

\bibitem{DBLP:conf/acsac/ArastehMRH24}
\BIBentryALTinterwordspacing
S.~Arasteh, J.~Mirkovic, M.~Raghothaman, and C.~Hauser, ``Binhunter: {A} fine-grained graph representation for localizing vulnerabilities in binary executables\({}^{\mbox{*}}\),'' in \emph{Annual Computer Security Applications Conference, {ACSAC} 2024, Honolulu, HI, USA, December 9-13, 2024}.\hskip 1em plus 0.5em minus 0.4em\relax {IEEE}, 2024, pp. 1062--1074. [Online]. Available: \url{https://doi.org/10.1109/ACSAC63791.2024.00087}
\BIBentrySTDinterwordspacing

\bibitem{DBLP:conf/uss/AndriesseCVSB16}
\BIBentryALTinterwordspacing
D.~Andriesse, X.~Chen, V.~van~der Veen, A.~Slowinska, and H.~Bos, ``An in-depth analysis of disassembly on full-scale x86/x64 binaries,'' in \emph{25th {USENIX} Security Symposium, {USENIX} Security 16, Austin, TX, USA, August 10-12, 2016}, T.~Holz and S.~Savage, Eds.\hskip 1em plus 0.5em minus 0.4em\relax {USENIX} Association, 2016, pp. 583--600. [Online]. Available: \url{https://www.usenix.org/conference/usenixsecurity16/technical-sessions/presentation/andriesse}
\BIBentrySTDinterwordspacing

\bibitem{DBLP:conf/uss/PangZYM022}
\BIBentryALTinterwordspacing
C.~Pang, T.~Zhang, R.~Yu, B.~Mao, and J.~Xu, ``Ground truth for binary disassembly is not easy,'' in \emph{31st {USENIX} Security Symposium, {USENIX} Security 2022, Boston, MA, USA, August 10-12, 2022}, K.~R.~B. Butler and K.~Thomas, Eds.\hskip 1em plus 0.5em minus 0.4em\relax {USENIX} Association, 2022, pp. 2479--2495. [Online]. Available: \url{https://www.usenix.org/conference/usenixsecurity22/presentation/pang-chengbin}
\BIBentrySTDinterwordspacing

\bibitem{DBLP:conf/sp/PangYCKPMX21}
\BIBentryALTinterwordspacing
C.~Pang, R.~Yu, Y.~Chen, E.~Koskinen, G.~Portokalidis, B.~Mao, and J.~Xu, ``Sok: All you ever wanted to know about x86/x64 binary disassembly but were afraid to ask,'' in \emph{42nd {IEEE} Symposium on Security and Privacy, {SP} 2021, San Francisco, CA, USA, 24-27 May 2021}.\hskip 1em plus 0.5em minus 0.4em\relax {IEEE}, 2021, pp. 833--851. [Online]. Available: \url{https://doi.org/10.1109/SP40001.2021.00012}
\BIBentrySTDinterwordspacing

\bibitem{DBLP:conf/cc/FedericoPA17}
\BIBentryALTinterwordspacing
A.~D. Federico, M.~Payer, and G.~Agosta, ``rev.ng: a unified binary analysis framework to recover cfgs and function boundaries,'' in \emph{Proceedings of the 26th International Conference on Compiler Construction, Austin, TX, USA, February 5-6, 2017}, P.~Wu and S.~Hack, Eds.\hskip 1em plus 0.5em minus 0.4em\relax {ACM}, 2017, pp. 131--141. [Online]. Available: \url{http://dl.acm.org/citation.cfm?id=3033028}
\BIBentrySTDinterwordspacing

\bibitem{DBLP:conf/paste/Hind01}
\BIBentryALTinterwordspacing
M.~Hind, ``Pointer analysis: haven't we solved this problem yet?'' in \emph{Proceedings of the 2001 {ACM} {SIGPLAN-SIGSOFT} Workshop on Program Analysis For Software Tools and Engineering, PASTE'01, Snowbird, Utah, USA, June 18-19, 2001}, J.~Field and G.~Snelting, Eds.\hskip 1em plus 0.5em minus 0.4em\relax {ACM}, 2001, pp. 54--61. [Online]. Available: \url{https://doi.org/10.1145/379605.379665}
\BIBentrySTDinterwordspacing

\bibitem{autofacts}
\BIBentryALTinterwordspacing
A.~Machiry, N.~Redini, E.~Gustafson, H.~Aghakhani, C.~Kruegel, and G.~Vigna, ``Towards automatically generating a sound and complete dataset for evaluating static analysis tools,'' \emph{Workshop on Binary Analysis Research (BAR)}, 2019. [Online]. Available: \url{https://par.nsf.gov/biblio/10155111}
\BIBentrySTDinterwordspacing

\end{thebibliography}

\end{document}